\documentclass[twocolumn, usenatbib]{mnras}
\usepackage[utf8]{inputenc}
\usepackage{amssymb}
\usepackage{graphicx}
\usepackage{amsmath}
\usepackage{float}
\usepackage{subfigure}
\usepackage{hyperref}
\usepackage{dsfont}
\usepackage{balance}

\newcommand{\eps}{\boldsymbol{\varepsilon}}
\newcommand{\R}{\mathbb{R}}
\newcommand{\HH}{\mathcal{H}}

\title[Bayesian jackknife tests with a small number of subsets]{Bayesian jackknife tests with a small number of subsets: Application to HERA 21cm power spectrum upper limits \vspace{-0.8em}}

\author[M.\ J.\ Wilensky et al.]{\Large Michael J. Wilensky$^{1,2}$\thanks{Email: michael.wilensky@manchester.ac.uk},
	Fraser  Kennedy$^{2}$,
	Philip  Bull$^{1,3,2}$,
	Joshua S. Dillon$^{4}$,
	Zara  Abdurashidova$^{4}$,
\newauthor \Large
    Tyrone  Adams$^{5}$,
	James E. Aguirre$^{6}$,
	Paul  Alexander$^{7}$,
	Zaki S. Ali$^{4}$,
	Rushelle  Baartman$^{5}$,
	Yanga  Balfour$^{5}$,
\newauthor \Large
	Adam P. Beardsley$^{8,9}$,
	Gianni  Bernardi$^{10,11,5}$,
	Tashalee S. Billings$^{6}$,
	Judd D. Bowman$^{8}$,
\newauthor \Large
	Richard F. Bradley$^{12}$,
	Jacob  Burba$^{13}$,
	Steven Carey$^{7}$,
	Chris L. Carilli$^{14}$,
	Carina Cheng$^{4}$,
\newauthor \Large
	David R. DeBoer$^{15}$,
	Eloy  de~Lera~Acedo$^{7}$,
	Matt  Dexter$^{15}$,
	Nico  Eksteen$^{5}$,
\newauthor \Large
	John  Ely$^{7}$,
	Aaron  Ewall-Wice$^{4,16}$,
	Nicolas  Fagnoni$^{7}$,
	Randall  Fritz$^{5}$,
	Steven R. Furlanetto$^{17}$,
\newauthor \Large
	Kingsley  Gale-Sides$^{7}$,
	Brian  Glendenning$^{18}$,
	Deepthi  Gorthi$^{4}$,
	Bradley  Greig$^{19}$,
	Jasper  Grobbelaar$^{5}$,
\newauthor \Large
	Ziyaad  Halday$^{5}$,
	Bryna J. Hazelton$^{20,21}$,
	Jacqueline N. Hewitt$^{22,23}$,
	Jack  Hickish$^{15}$,
	Daniel C. Jacobs$^{8}$,
\newauthor \Large
	Austin Julius$^{5}$,
	MacCalvin  Kariseb$^{5}$,
	Nicholas S. Kern$^{4,23}$,
	Joshua  Kerrigan$^{13}$,
	Piyanat  Kittiwisit$^{3}$,
\newauthor \Large
	Saul A. Kohn$^{6}$,
	Matthew  Kolopanis$^{8}$,
	Adam  Lanman$^{13}$,
	Paul  La~Plante$^{4,6}$,
	Adrian  Liu$^{24}$,
\newauthor \Large
	Anita Loots$^{5}$,
	David Harold~Edward MacMahon$^{15}$,
	Lourence  Malan$^{5}$,
	Cresshim  Malgas$^{5}$,
\newauthor \Large
	Keith  Malgas$^{5}$,
	Bradley  Marero$^{5}$,
	Zachary E. Martinot$^{6}$,
	Andrei  Mesinger$^{25}$,
	Mathakane  Molewa$^{5}$,
\newauthor \Large
	Miguel F. Morales$^{20}$,
	Tshegofalang  Mosiane$^{5}$,
	Steven G. Murray$^{8}$,
	Abraham R. Neben$^{23}$,
\newauthor \Large
	Bojan  Nikolic$^{7}$,
	Hans  Nuwegeld$^{5}$,
	Aaron R. Parsons$^{4}$,
	Nipanjana  Patra$^{4}$,
	Samantha  Pieterse$^{5}$,
\newauthor \Large
	Nima  Razavi-Ghods$^{7}$,
	James  Robnett$^{14}$,
	Kathryn  Rosie$^{5}$,
	Peter  Sims$^{24}$,
\newauthor \Large
	Hilton  Swarts$^{5}$,
	Nithyanandan  Thyagarajan$^{26,14}$,
	Pieter  van~Wyngaarden$^{5}$,
\newauthor \Large
	Peter K.~G. Williams$^{27,28}$,
	Haoxuan  Zheng$^{23}$
\vspace{0.5em} \\ 
$^{1}$ Jodrell Bank Centre for Astrophysics, University of Manchester, Manchester M13 9PL, UK\\
$^{2}$ Astronomy Unit, Queen Mary University of London, London E1 4NS, UK\\
$^{3}$ Department of Physics and Astronomy,  University of Western Cape, Cape Town, 7535, South Africa\\
$^{4}$ Department of Astronomy, University of California, Berkeley, CA\\
$^{5}$ South African Radio Astronomy Observatory, Black River Park, 2 Fir Street, Observatory, Cape Town, 7925, South Africa\\
$^{6}$ Department of Physics and Astronomy, University of Pennsylvania, Philadelphia, PA\\
$^{7}$ Cavendish Astrophysics, University of Cambridge, Cambridge, UK\\
$^{8}$ School of Earth and Space Exploration, Arizona State University, Tempe, AZ\\
$^{9}$ Department of Physics, Winona State University, Winona, MN\\
$^{\dagger}$ NSF Astronomy and Astrophysics Postdoctoral Fellow\\
$^{10}$ INAF-Istituto di Radioastronomia, via Gobetti 101, 40129 Bologna, Italy\\
$^{11}$ Department of Physics and Electronics, Rhodes University, PO Box 94, Grahamstown, 6140, South Africa\\
$^{12}$ National Radio Astronomy Observatory, Charlottesville, VA\\
$^{13}$ Department of Physics, Brown University, Providence, RI\\
$^{14}$ National Radio Astronomy Observatory, Socorro, NM 87801, USA\\
$^{15}$ Radio Astronomy Lab, University of California, Berkeley, CA\\
$^{16}$ Department of Physics, University of California, Berkeley, CA\\
$^{17}$ Department of Physics and Astronomy, University of California, Los Angeles, CA\\
$^{18}$ National Radio Astronomy Observatory, Socorro, NM\\
$^{19}$ School of Physics, University of Melbourne, Parkville, VIC 3010, Australia\\
$^{20}$ Department of Physics, University of Washington, Seattle, WA\\
$^{21}$ eScience Institute, University of Washington, Seattle, WA\\
$^{22}$ MIT Kavli Institute, Massachusetts Institute of Technology, Cambridge, MA\\
$^{23}$ Department of Physics, Massachusetts Institute of Technology, Cambridge, MA\\
$^{24}$ Department of Physics and McGill Space Institute, McGill University, 3600 University Street, Montreal, QC H3A 2T8, Canada\\
$^{25}$ Scuola Normale Superiore, 56126 Pisa, PI, Italy\\
$^{26}$ Commonwealth Scientific and Industrial Research Organisation (CSIRO), Space \& Astronomy, P. O. Box 1130, Bentley, WA 6102, Australia\\
$^{27}$ Center for Astrophysics, Harvard \& Smithsonian, Cambridge, MA\\
$^{28}$ American Astronomical Society, Washington, DC
}

\date{31 October 2022}

\begin{document}
\maketitle

\begin{abstract}
    We present a Bayesian jackknife test for assessing the probability that a data set contains biased subsets, and, if so, which of the subsets are likely to be biased. The test can be used to assess the presence and likely source of statistical tension between different measurements of the same quantities in an automated manner.
    Under certain broadly applicable assumptions, the test is analytically tractable. We also provide an open source code, \href{https://github.com/mwilensky768/chiborg}{\textsc{chiborg}}, that performs both analytic and numerical computations of the test on general Gaussian-distributed data. After exploring the information theoretical aspects of the test and its performance with an array of simulations, we apply it to data from the Hydrogen Epoch of Reionization Array (HERA) to assess whether different sub-seasons of observing can justifiably be combined to produce a deeper 21cm power spectrum upper limit. We find that, with a handful of exceptions, the HERA data in question are statistically consistent and this decision is justified. We conclude by pointing out the wide applicability of this test, including to CMB experiments and the $H_0$ tension.
\end{abstract}


\section{Introduction}

Evaluating the presence and significance of statistical tension is a core part of the process of reconciling independent measurements of the same physical quantity. A lack of tension permits us to combine measurements in order to place improved constraints on quantities of interest, while the presence of tension raises the prospect of contamination of some or all of the data by previously unidentified systematic effects, or new physical effects not anticipated in the theoretical model being used to interpret the data.

A topical example within the cosmology community is the `Hubble tension' \citep{2013PDU.....2..166V, 2020PhRvD.101d3533K, 2021ApJ...908L...6R, 2021CQGra..38o3001D, Freedman2021}, which arises from apparent inconsistencies in the measured value of the Hubble parameter, $H_0$, obtained from different observational probes. In this case, the fact that there is a tension between at least some of the measurements is not in dispute; a difference of almost $5\sigma$ is found between the reported values of $H_0$ from the Planck CMB mission and local distance ladder measurements for example \citep{2021ApJ...908L...6R}. The question instead is which of the points are discrepant, and why. By finding distinct subsets of the measurements that are internally statistically consistent (and so can be combined to produce a more precise measurement), but in tension with other subsets, we can hope to identify a systematic bias or physical phenomenon that explains the tension \citep[e.g.][]{2021CQGra..38o3001D, Blanchard2022}. Similar tensions between other cosmological parameters have also been tentatively identified and studied \citep[e.g.][]{2017PhRvD..96h3532L, Raveri2019, 2021MNRAS.505.6179L}.

Assessing the probability that a given measurement is discrepant is difficult (and potentially ill-defined) without some knowledge of the expected distribution of the measurements and possible biases, particularly when only a handful of measurements are available. A Bayesian approach lends itself particularly well to this kind of question, as it allows us to set out the statistical questions we are asking of the data in an explicit manner, e.g. by requiring definite statements of the models/hypotheses and prior assumptions. A wide variety of Bayesian approaches have previously been applied to cosmological tensions \citep[e.g.][]{1997upa..conf...49P, 2007MNRAS.377L..74L, 2013PDU.....2..166V, 2015MNRAS.451.2610H, 2016A&A...588A..19L, 2016PhRvD..93j3507S, 2017PhRvD..96h3532L, Raveri2019, 2019PhRvD.100d3504H, Blanchard2022}, and tend to share some common features. For instance, many employ model comparison tests between a null hypothesis and a variety of alternative model scenarios, integrating over a chosen prior (with an appropriate penalty/`Occam factor' to penalise more flexible models) to determine which scenario has overall higher odds given the available data \citep[e.g.][]{2007MNRAS.377L..74L, 2007MNRAS.378...72T}. Some also employ methods that measure a `distance' between the empirical distribution of the measured values and the expected/ideal distribution \citep[e.g.][]{2016A&A...588A..19L, 2019JCAP...01..011N}, but choose different ways of quantifying the distance and determining what counts as a significant difference. These latter two points, plus different strategies for choosing (or even disregarding) priors, account for much of the variety in these approaches.

Most of the references that we have included above focus on assessing tensions between the values of cosmological parameters inferred from different experiments. Another common situation that requires a similar type of assessment of statistical consistency is the combination of different subsets or `seasons' of data from a single experiment. Particularly in CMB and radio (21cm intensity mapping) experiments, long integration times are required to reach sufficient signal-to-noise on the faint signals that are being targeted. This often requires observations that stretch over multiple observing seasons, taken across several years and different times of the year, and possibly even different instrument configurations etc. The potential for discrepancies due to varying systematic effects between subsets of observations is therefore relatively high, and so one is usually interested in performing `jackknife' tests that attempt to determine whether the subsets are consistent with one another. Assuming that they are, the data can then be combined in order to improve sensitivity. Otherwise, the discrepancy can be taken as evidence of systematic contamination that must be addressed.

We note that Bayesian approaches seem to be less commonly used for this kind of assessment however. Instead, frequentist simulation-based methods seem to be more prevalent, for instance ones in which many simulations of a null hypothesis are passed through the same data analysis pipeline and compared at the level of the measured statistic (e.g. a power spectrum). The measured value of each data point (e.g. each power spectrum bandpower) is then compared with the distribution of simulated null hypothesis measurements to calculate a $p$-value or ``Probability-to-Exceed'' (PTE) \citep[e.g.][]{2020A&A...641A...7P, 2022ApJ...927...77A, 2022PhRvD.105b3520A}, which gives an estimate of whether each data point can be considered an `outlier' or not, subject to a potentially large list of implicit assumptions. Part of the reason for the popularity of this approach is likely to be the relative complexity of the `forward model' of the data analysis process, which can involve many non-trivial steps, hence lending itself to a simulation-based method. There is also the question of whether a more careful (and somewhat more cumbersome) Bayesian approach is overly complex when only approximate agreement between subsets of data is typically considered sufficient to proceed with averaging them together.

Certain types of observation, such as CMB B-mode searches and 21cm intensity mapping, now routinely involve data with very large dynamic ranges, for which a more controlled assessment of statistical consistency may ultimately be required in order to recover the extremely faint target signals from combinations of observing seasons. 
In this work, we develop a Bayesian jackknife test that has such an application in mind. The test is constructed to determine whether any subsets of the data are biased, and if so, which subsets. The general framework is flexible in that the analyst may sculpt the bias hypotheses by choosing appropriate priors, and is not limited to a particular set of bias configurations. We implement this framework in an open source \textsc{Python} software package \href{https://github.com/mwilensky768/chiborg}{\textsc{chiborg}},\footnote{\url{https://github.com/mwilensky768/chiborg}} named as an alternative to ``$\chi$-by-eye'' approaches to similar problems. In essence, the test computes the evidence (marginal likelihood) for each bias hypothesis, and combines these into a posterior probability distribution (PPD). The most probable hypothesis can then be identified, or, depending on the application, a loss function may be minimized over the posterior to decide on the outcome of the test. While similar approaches have been developed before, e.g. to assess cosmological parameter tensions, we place an additional focus on the decision-making aspect of the test, pointing out how data analysts can use it to select or exclude subsets of the data for further averaging, as well as to hunt down systematic effects.

Alternative Bayesian approaches to our statistical consistency checks include fitting a hierarchical model where (for example) the explanatory parameter is a covariance parameter between the underlying means of the data subsets,\footnote{In some respects this resembles the analysis suggested in this work, but we have elected to only consider a discrete selection of covariance models whereas a typical hierarchical fitting procedure would allow this parameter to vary continuously.} and/or performing posterior predictive checks to see if a model inferred from a subset of the data can produce simulated data sets that are consistent with the remaining measured data (the subset not used for the inference). Both of these procedures are introduced and explored in a broader statistical context in \citet{gelmanbda04}, which reviews Bayesian data analysis in general. Hierarchical modeling is more specifically explored in \citet{Gelman2009} and \citet{Isi2022}, while \citet{Rubin1984}, \citet{Meng1994}, \citet{Gelman1996}, \citet{2021MNRAS.503.2688D}, and \citet{Moran2022} focus on posterior predictive methods. In posterior predictive checks, one simulates fictitious data by sampling its PPD as inferred from (a subset of) the data, and then exposes the original data to a well-defined consistency test within the simulated data such as a $\chi^2$ test (whose sampling distribution can be built up from the ensemble of simulated data). They are appealing since they are well-formulated and clearly assess the question of statistical consistency. However, implementations that focus on the posterior predictive $p$-value suffer from the need to arbitrate statistical significance, i.e. it is arbitrary when to accept or reject the hypothesis of statistical consistency. We address this by integrating our inference into a decision-making framework. The relative consistency between the data subsets is represented by a set of hypotheses each with an associated Bayesian posterior. These posteriors can be weighed against a loss function based on inference error in order to determine the optimal conclusion for the needs at hand. In a hierarchical modeling scheme, one could in principle perform the same exercise after obtaining the posterior for the explanatory parameter. Another important strength of our jackknife test is that it assesses the question of statistical consistency by pointing directly to the problematic data subsets in an easily interpreted fashion. In other words, once the data are known to contain inconsistencies, the next most immediate question is often ``which data are inconsistent?"

We have in mind a particular scenario in the analysis of 21cm array data from the Hydrogen Epoch of Reionization Array (HERA) in \cite{HERA2022b}, which motivated the design of this test, and its application when only small numbers of subsets of data are available (so that the distribution of the subsets is difficult to determine empirically). Specifically, we use the test in order to develop a (post hoc) statistical justification for coherently combining several disjoint subsets of data into one data set for the purposes of placing an improved upper limit on the cosmological 21cm power spectrum signal. We explain this in more detail in what follows. For specific reviews on the subject of 21cm cosmology, we direct the reader to \citet{Furl2006}, \citet{Morales2010}, and \citet{Liu2020}, among others.

This paper is laid out as follows. In \S\ref{sec:math}, we lay down a theoretical foundation for the test. We also explain from an information theoretical standpoint how to design the test for different applications. We then apply the test to a suite of simple simulations in \S\ref{sec:toy} in order to explore the performance and behavior of the test in different circumstances. In \S\ref{sec:HERA_data}, we use the test on simulated HERA power spectra using the HERA validation pipeline \citep{Aguirre2022}, as well as the actual power spectrum measurements from \citet{HERA2022b}. Finally, we provide a summary of the work and our conclusions in \S\ref{sec:conc}.

\section{Mathematical Formalism}
\label{sec:math}

In this section we establish the mathematical formalism for detecting biases in data. We begin by stating the problem and defining our test in terms of a null hypothesis and alternate hypotheses, stating what each of these means mathematically. We then analytically derive the posterior probabilities of these hypotheses given some data.

\subsection{Problem Statement}

We consider the general problem of inferring whether a set of normally distributed data points are drawn about the same mean. We have in mind a particular choice of data in the HERA power spectrum estimation pipeline, and use this example in order to guide the discussion with a concrete scenario. However, the formalism could be applied in other data analysis problems, such as determining whether different measurements of the Hubble parameter are statistically consistent. 

Consider a scenario in which the HERA analysts have carefully prepared data for the estimation of the 21cm power spectrum signal, and have chosen to separate the data into $N$ observing epochs, all disjoint (e.g. 20 nights, separated into $N=4$ disjoint sets of 5 nights), and expected to be similar in quality. For the final power spectrum limits, these epochs will be combined into one measurement. However, as an auxiliary data set, they form power spectrum estimates for each epoch independently. We may then ask the question, ``given these power spectrum estimates, do we still believe these epochs are similar in character?" 

Clearly we must be more specific since there are many ways in which the data can be inconsistent with one another. In the case of 21cm power spectrum estimation, we are often concerned with whether there is a systematic bias. The framework presented in this paper is flexible in that it can answer different statistical questions about the relationships of the potential biases based on which hypotheses are considered. For instance, one question that it can answer is, ``which subsets of the data are likely to possess a significant bias of unknown strength?" Another is, ``given a particular strength of bias that can affect some data but not others (for whatever the reason), how likely is it that this bias is present?" Yet another is, ``given that we believe some subsets of the data are biased, which partitions of these subsets have biases that are similar to one another?"

In the following subsections, we outline a Bayesian hypothesis test to answer these questions. We then provide a formulation of the marginal likelihoods corresponding to a general set of relationships between potential biases of unknown strength. From these likelihoods and corresponding priors, we calculate the posterior odds of various hypothetical bias relationships against one another, allowing us to form a conclusion about the nature of the biases within the data. 

\subsection{Phrasing the Hypothesis Test}

We are given $N$ data points arranged into a vector, $\bold{d}$, and suppose that each datum may be biased by some unknown value, $\eps$. In the absence of a bias, each datum is drawn about the same mean value, $\mu_0$ (potentially unknown), according a Gaussian distribution with known covariance $\bold{C}_0$. We define this as our null hypothesis, $\HH_0$. In the HERA example, the data would be the bandpower estimates for a single wave mode, and $\bold{C}_0$ would be a diagonal matrix where each entry would be the estimated variance of the corresponding bandpower.

In the HERA problem, there are multiple choices for the interpretation of $\mu_0$. For example, one could choose a value from which systematics are expected to be subtracted from theoretical work (such as the validation pipeline accompanying the data analysis software), a chosen reionization model to be investigated, or even the value 0. In this final case, a rejection of the null hypothesis for the real component of the HERA bandpower is ambivalent between whether it is a detection of the reionization signal or some systematic causing the bias, and one would need to program specific hypotheses about the bias (or perform an independent analysis) in order to distinguish between the two.

In general, the scale of the problem is set by the size of the error bars, and we observe that differences in choice of $\mu_0$ that are much smaller than the statistical error associated with the data make imperceptible differences in the test results. This has potential consequences regarding a choice of prior on $\mu_0$, explained in \S\ref{sec:parameter_choice}.

To answer a particular statistical question about the biases, we formulate hypotheses about them, and compare the posterior probabilities among one another using the ``odds" of one hypothesis to another, conditional on the data and known covariance. Mathematically, we write,
\begin{equation}
    P(\HH_i | \bold{d}, \bold{C}_0) = \frac{P(\bold{d} | \HH_i, \bold{C}_0)P(\HH_i | \bold{C}_0)}{P(\bold{d} | \bold{C}_0)}.
    \label{eq:bayes}
\end{equation}
where $\HH_i$ is shorthand for a specification of knowledge about the biases, formally detailed in the next subsection. In the formalism in this paper, we do not use our knowledge of the data covariance to inform the prior of the hypothesis, though we imagine that there can be good reasons for doing otherwise. This allows us to write
\begin{equation}
    P(\mathcal{H}_i | \bold{C}_0) = P(\mathcal{H}_i),
\end{equation}
If there are $Q$ hypotheses in consideration (including the null hypothesis), then the denominator of Eq.~\ref{eq:bayes} is simply calculated as
\begin{equation}
    P(\bold{d} | \bold{C}_0) = \sum_{i=0}^{Q - 1} P(\bold{d} | \mathcal{H}_i, \bold{C}_0)P(\mathcal{H}_i).
    \label{eq:total_evid}
\end{equation}

\subsection{Evaluating the Posterior}
\label{sec:post_eval}

In order to evaluate the posterior probability of any hypothesis, we must handle the nuisance parameters by marginalizing over corresponding priors in order to produce the likelihood. For instance, if we expected tbe data to possess a common mean that takes the value $\mu_0$, we would say
\begin{equation}
    P(\bold{d} | \bold{C}_0, \mu_0, \mathcal{H}_0) = \frac{\exp \big(-\frac{1}{2}(\bold{d} - \mu_0\mathds{1})^T\bold{C}_0^{-1}(\bold{d} - \mu_0\mathds{1})\big)}{\sqrt{\det(2\pi\bold{C}_0)}}.
\end{equation}
where $\mathds{1}$ is a vector for which every component is equal to 1. Similarly, if we knew the data to be drawn from a biased Gaussian, with bias given by the vector $\eps$, we would write
\begin{multline}
    P(\bold{d} | \bold{C}_0, \mu_0, \eps, \mathcal{H}_i) = \\
    \frac{\exp \big(-\frac{1}{2}(\bold{d} - \mu_0\mathds{1} - \eps)^T\bold{C}_0^{-1}(\bold{d} - \mu_0\mathds{1} - \eps)\big)}{\sqrt{\det(2\pi\bold{C}_0)}}.
\end{multline}
In general, we need not assume particular values for the test. We instead propose that the means and biases may take a range of values, with weights assigned by prior probabilities:
\begin{equation}
    P(\bold{d} | \bold{C}_0, \mathcal{H}_0) = \int_\R d\mu_0P(\mu_0)P(\bold{d} | \bold{C}_0, \mu_0, \mathcal{H}_0)
    \label{eq:H0_marg}
\end{equation}
\begin{equation}
    P(\bold{d} | \bold{C}_0, \mathcal{H}_i) = \int_\R d\mu_0P(\mu_0)\int_{\R^N}d\eps P(\eps | \HH_i)P(\bold{d} | \bold{C}_0, \mu_0, \eps, \mathcal{H}_i)
    \label{eq:H1_marg}
\end{equation}
where $P(\mu_0)$ and $P(\eps | \HH_i)$ are suitable priors that reflect a combination of our knowledge about the system and in some part the nature of the types of biases we would like to inspect. We have implicitly assumed that $P(\mu_0)$ is the same regardless of which hypothesis is true, since the change in that parameter under a violation of the null hypothesis is explicitly parameterized by $\eps$. 

If one chooses Gaussian priors for these nuisance parameters, then the calculation has a closed-form solution. We acknowledge that in some instances this choice of prior may seem inappropriate. In the HERA example at hand, if $\mu_0$ is meant to represent the value of the 21cm EoR power spectrum signal for a particular wave mode, then assigning nonzero probability to a negative value would be incorrect in a very strict treatment. If this value is well-known, then the width of the prior might be such that negative values are negligibly probable anyway. However, the current state of the field is that this value is not particularly well-constrained for any wave mode. Fortunately, this is a 1-dimensional integral, and so reasonably arbitrary positive-definite priors can be marginalized over numerically. There are also a number of positive-definite priors with shape and scale parameters that can yield a closed-form solution when integrated against the Gaussian likelihood, should this be the chosen interpretation for $\mu_0$. 
We show the derivation using Gaussian priors. 

An alternative interpretation for $\mu_0$ in the HERA context when applying this result is that it is a common bias which we expect each epoch to be limited by e.g. the expected strength of residual cable reflections or radio frequency interference (RFI) based on our understanding of systematic mitigation. An alternative hypothesis then reflects an instance where systematic mitigation was not as effective as expected in one or more epochs (or potentially far stronger than expected in the case of a negative bias, indicating signal loss). 

We explicitly write our priors like so:
\begin{equation}
    P(\mu_0) = \frac{\exp\big(-(\mu_0 - \mu_p)^2/\sigma_p^2\big)}{\sqrt{2\pi\sigma_p^2}}
\end{equation}
\begin{equation}
    P(\eps | \HH_i) = \frac{\exp \big(-\frac{1}{2}(\eps - \boldsymbol{\mu}_{\eps, i})^T\bold{C}_{\eps, i}^{-1}(\eps - \boldsymbol{\mu}_{\eps, i}\big)}{\sqrt{\det(2\pi\bold{C}_{\eps, i})}},
\end{equation}
where we allow for potential correlations in the bias vector by way of $\bold{C}_{\eps, i}$. In other words, each alternative hypothesis, $\HH_i$, is represented by a mean value, and a relationship between variations in the bias parameters. Having these degrees of freedom allows the analyst to sculpt the statistical question. For instance, taking the limit as $\bold{C}_{\eps, i}$ goes to the $\bold{0}$ matrix produces a delta function in the marginalization, which is equivalent to conditioning on a bias equal to $\bmu_{\eps,i}$. As another example, if one already believes that all the data are biased, then one may compare a hypothesis with diagonal covariance against one with a highly degenerate covariance, where the degeneracies probe exact (or nearly exact) equality relationships between the biases. We discuss this in more detail in \S\ref{sec:parameter_choice}.


\begin{figure*}
    \centering
    \includegraphics[width=\linewidth]{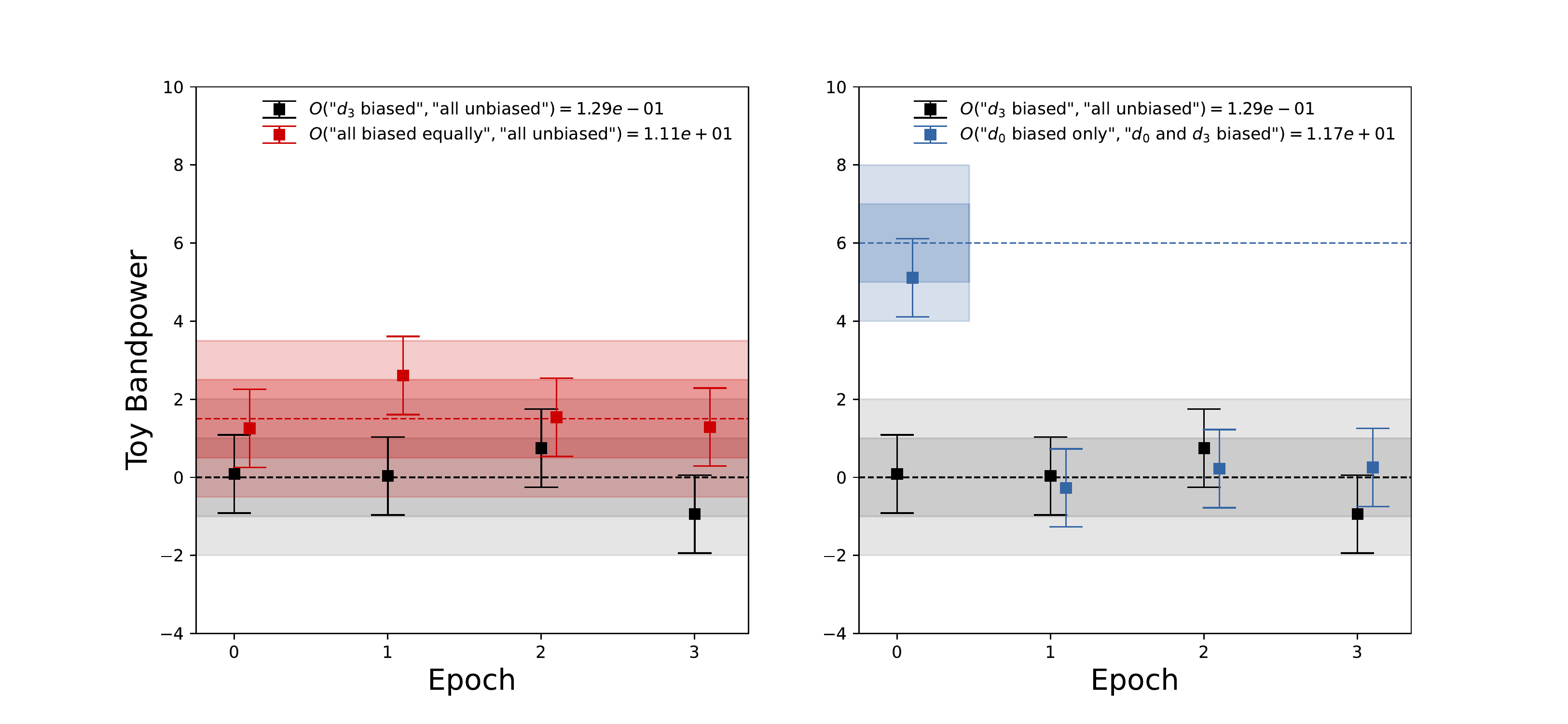}
    
    \caption{Toy representation of the type of problem that is solvable by the formalism in this paper. Depicted are several pseudorandom draws of normally distributed random variables, color coded according to their mean (dashed), and their 1-and-2$\sigma$ confidence interval (shaded). Definitively concluding whether there is a significant bias in any given data set by eye (doing a $\chi$-by-eye) is difficult in some circumstances. By formalizing different bias scenarios as Bayesian hypotheses, we are able to produce the odds of different scenarios given the data (legend). Even using very broad priors about how strong the biases could be, we are able to resolve each correct hypothesis against its nearest competitor. By using more informative bias priors (which results in more specific versions of each hypothesis), we can bring the odds listed for the red data over 100, and the odds listed for the blue data to an extremely high exponent.}
    \label{fig:int_builder}
\end{figure*}

Since Equation \ref{eq:H0_marg} is a limiting case of Equation \ref{eq:H1_marg}, we focus on Equation \ref{eq:H1_marg}. The integral over $\eps$ is a multidimensional Gaussian convolution. The result, which can be derived by completing the square and some matrix manipulation (or the convolution theorem), is essentially identical to the 1-dimensional case. We obtain
\begin{multline}
    f(\mu_0) \equiv \int_{\R^N}d\eps P(\eps)P(\bold{d} | \bold{C}_0, \mu_0, \eps, \HH_i) \\
     = \frac{\exp \big(-\frac{1}{2}(\bold{d} - \mu_0\mathds{1} - \bmu_{\eps, i} )^T(\bold{C}_{\eps, i} + \bold{C}_0)^{-1}(\bold{d} - \mu_0\mathds{1} - \bmu_{\eps, i})\big)}{\sqrt{\det(2\pi(\bold{C}_{\eps, i} + \bold{C}_0))}} \nonumber.
\end{multline}
We must then integrate this function, which is Gaussian in $\mu_0$, against the $\mu_0$ prior. However, it is not a probability density function in $\mu_0$, so we must be a little careful. By inspection of the quadratic term, the ``variance" of this Gaussian function, $\tilde{\sigma}_i^2$, is defined by
\begin{equation}
    \frac{1}{\tilde{\sigma}_i^2} = \mathds{1}^T(\bold{C}_{\eps, i} + \bold{C}_0)^{-1}\mathds{1}.
    \label{eq:sig_tild}
\end{equation}
To see the ``mean" of this Gaussian function, $\tilde{\mu}_i$ we do a standard maximization procedure:
\begin{equation}
    \frac{d\ln f}{d\mu_0} \bigg|_{\tilde{\mu}_i} = \mathds{1}^T(\bold{C}_{\eps, i} + \bold{C}_0)^{-1}(\bold{d} - \tilde{\mu}_i\mathds{1} - \bmu_{\eps, i}) = 0.
\end{equation}
This is satisfied when
\begin{equation}
    \tilde{\mu}_i = \tilde{\sigma}_i^2\mathds{1}^T(\bold{C}_{\eps, i} + \bold{C}_0)^{-1}(\bold{d} - \bmu_{\eps, i}).
    \label{eq:mu_tild}
\end{equation}
Completing the square and performing the integral over $\mu_0$ yields
\begin{multline}
    P(\bold{d} | \bold{C}_0, \HH_i) = \frac{\exp\big(-\frac{1}{2}\big(\tilde{\mu}_i - \mu_p\big)^2/(\sigma_p^2 + \tilde{\sigma}_i^2)\big)}{\sqrt{2\pi(\tilde{\sigma}_i^2 + \sigma_p^2)}} \\
    \times {\exp\big(-\frac{1}{2}[(\bold{d} - \bmu_{\eps, i})^T(\bold{C}_0+\bold{C}_{\eps, i})^{-1}(\bold{d} - \bmu_{\eps, i}) - \tilde{\mu}_i^2/\tilde{\sigma}_i^2]\big)}\\
    \times \frac{\sqrt{2\pi\tilde{\sigma}_i^2}}{\sqrt{\det(2\pi(\bold{C}_0 + \bold{C}_{\eps, i}))}}.
    \label{eq:post_1}
\end{multline}

Since this is a probability distribution in $\bold{d}$, and is of Gaussian form, it can be written
\begin{equation}
    P(\bold{d} | \bold{C}_0, \HH_i) = \frac{\exp \big(-\frac{1}{2}(\bold{d} - \bmu'_i)^T\bold{C}'^{-1}_i(\bold{d} - \bmu'_i)\big)}{\sqrt{\det(2\pi\bold{C}'_i)}}.
\end{equation}
In one final exercise in completing the square, one finds that
\begin{equation}
    \bmu'_i = \bmu_{\eps} + \mu_p\mathds{1}
    \label{eq:final_mean}
\end{equation}
\begin{equation}
    \bold{C}'^{-1}_i = (\bold{C}_0 + \bold{C}_{\eps, i})^{-1} - \frac{(\bold{C}_0 + \bold{C}_{\eps, i})^{-1}\mathds{1}\mathds{1}^T(\bold{C}_0 + \bold{C}_{\eps, i})^{-1}}{{\tilde{\sigma}_i^{-2}} + {\sigma_p^{-2}}}
    \label{eq:final_cov}
\end{equation}

Now that the likelihoods are calculated, all that is necessary to produce the posterior probabilities is to multiply through by $P(\mathcal{H})$ and normalize by the total evidence (Equation \ref{eq:total_evid}). We explore decision rules in more detail in \S\ref{sec:toy}, but the simplest decision rule from these posteriors is to choose the maximum a posteriori hypothesis as the one `preferred' by the data.

We also make some remarks about the form of the likelihood. A particularly interesting quantity to consider is the logarithm of the odds of two hypotheses:
\begin{equation}
    \log O(\HH_i, \HH_j) = \log \left ( \frac{P(\HH_i | \bold{d}, \bold{C}_0)}{P(\HH_j | \bold{d}, \bold{C}_0)} \right )
\end{equation}
If one conditions on a particular mean and bias and uses a flat prior on $\{\HH_i\}$, then this quantity reduces to a difference in $\chi^2$ statistics with two different assumed models -- a familiar model comparison procedure from frequentist statistics. We can therefore think of comparing hypotheses this way as a Bayesian generalization to comparing the $\chi^2$ statistic of the data in the face of different models. In other words, this is the quantified version of the ``informed $\chi$-by-eye'' encouraged in the example we plot in Figure \ref{fig:int_builder}.

\subsection{Choosing Parameters for the Test}
\label{sec:parameter_choice}

In this section, we discuss parameter choices for each part of the test, which includes choice of prior probabilities, hyperparameters (parameters describing the priors) for continuous priors, and odds thresholds. We begin by discussing the choice of prior for $\mu_0$ in terms of the HERA problem.

\subsubsection{Choosing the prior for $\mu_0$}

We break the HERA problem into three scenarios and provide suggestions for each one. We note that when the measurement is significantly noise-dominated, the prior on $\mu_0$ does not drive the outcome of the inference provided it is less than the noise level.

\textbf{Residual systematics dominate the signal:} In this case, a Gaussian prior may be appropriate. For nonzero $\mu_0$, the interpretation would be that $\mu_0$ represents a common value to which the epochs are expected to be biased by systematics.  Negative values for some modes may represent large amount of signal loss. One could also condition on $\mu_0 = 0$, in which case a failure of the test would indicate the presence of significant systematics.

\textbf{Signal dominates the systematics, but is poorly constrained:} In this case, a Gaussian prior can be inappropriate due to the true signal power being strictly positive. There are at least two options. One option is to implement numerical integration techniques that can handle a non-Gaussian prior. Another option is to instead condition on $\mu_0 = 0$ and use a hypothesis for testing where all epochs are biased identically (or nearly so). In this scenario, ruling the data as all biased identically indicates that something signal-like has been located that is consistent between the epochs. Only further examination of the data can tell whether this is likely to be the signal of interest or some other, unexpected systematic. 

\textbf{Signal dominates the systematics, and is well constrained:} In this case we can use a Gaussian prior, where the width represents the strength of the constraint. The bias prior mean and/or width would need to be larger than this constraint and the error bar for the test to provide statistically meaningful results. A test failure would indicate an extreme malfunction in systematic mitigation. This is functionally the same as when systematics dominate the signal. All that changes is the meaning of the nuisance parameter.

We provide an open-source \textsc{Python} code called \href{https://github.com/mwilensky768/chiborg}{\textsc{chiborg}} for calculation of the posterior using analytically tractable priors as well as those that require numerical computation. This software is complete with documentation and example notebooks that explore various input and output options of the test to help users learn how to formulate the test for their purposes.

\subsubsection{Choosing the Bias Priors for Alternate Hypotheses}

The aim of this section is to define the subspace of bias hyperparameters that produce a well-defined hypothesis test using the information theoretical concept of ``mutual information." This requires concepts from information theory such as entropy of random variables, which we briefly review in Appendix~\ref{app:entropy}. In short, we wish to restrict our attention to scenarios (characterized by Gaussian bias priors with particular means and variances) that can be meaningfully distinguished from the null hypothesis and each other. Some biases will be too weak to distinguish from expected noise fluctuations around the null hypothesis, in which case our test will be inconclusive. By first finding the range of bias scenarios that are in principle distinguishable, we can substantially `sharpen' our test by only considering situations where a decisive result is possible.

In the following, we use the {\it mutual information} to determine the range of effects that are distinguishable using our hypothesis test. The basic idea is that we can obtain at most $S(P(\mathcal{H}))$ bits of information from the hypothesis test, where $S(P(x))$ is the Shannon entropy of a random variate $x$ (we may also write $S(x)$). Taking a two-hypothesis test with a flat prior as an example, this equates to (Equation \ref{eq:bin_ent}) $S_2(1/2) = 1$ bit of information. With 1 bit of information, we would be able to exactly specify the state of the random variable, $\HH$.\footnote{Said another way, to communicate the state of the random variable $\HH$, we require the full use of 1 bit on average. See Appendix \ref{app:entropy}.} For some data, specifically those where strong biases are common, the mutual information is close to maximal. This means that the average information provided by each application of the hypothesis test is sufficient to distinguish between data that come from the various hypotheses. Conversely, it is difficult to distinguish between a weak bias and what may just be a noise fluctuation; much less information is available in this case. Note that since the actual data observed by the analyst may deviate from the priors, we should be wary of saying that the information gained in practice is necessarily reflected by this relationship. We use this only as a rough guide for determining which priors are actually distinguishable with the volume of data used in the test.

\begin{figure}
    \centering
    \includegraphics[scale=0.65]{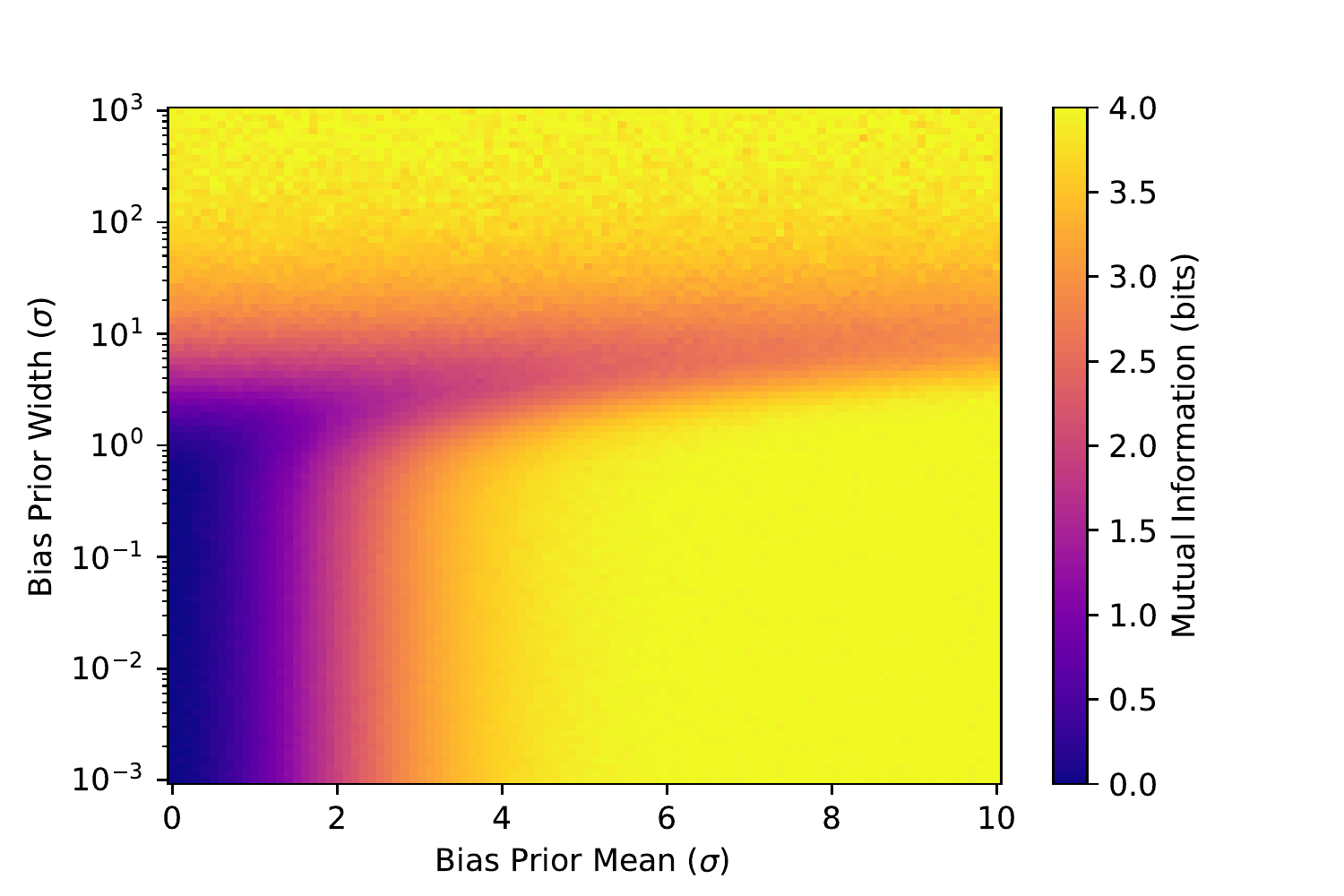}
    \caption{The mutual information provided by a hypothetical data set drawn according to the priors, as a function of bias prior width and mean (in units of the noise standard deviation, $\sigma$). The mutual information represents the reduction in uncertainty between two random variables when one is known -- in this case the data and the hypotheses (Appendix \ref{app:entropy}). Each draw has 4 data points (4 epochs in the HERA example), and we consider all 16 basic bias configurations (see main text), leading to a total of 4 bits that are required to specify the state of the system. We assume an exactly zero-mean null hypothesis. Small fluctuations are visible due to Monte Carlo estimation error, whose size is proportional to the bias prior width, hence the increased estimator noise at the top of the plot. In general, we can use this as a landscape for which hypotheses are distinguishable from one another. Bias priors corresponding to the bottom-left sector of the plot produce data that are only weakly biased. When these biases are approximately the size of the statistical error of the data or smaller, the test is, on average, unable to distinguish between any hypotheses. If biases are concentrated away from the origin, some confusion can arise if the bias prior is wide enough to have significant density at the origin. }
    \label{fig:mut_info}
\end{figure}

In Figure \ref{fig:mut_info}, we display the mutual information between the various test hypotheses and hypothetical data sets drawn according to the bias priors reflected by the choice of mean and width shown on the axes. We assume that each run of the test only has access to 4 data points, and condition on the underlying mean being equal to 0. Our hypothesis set can be represented by sixteen 4-bit strings, where a 1 in the $j$th entry of the string indicates that the $j$th datum is biased, and a 0 indicates it is unbiased. The corresponding bias prior is formulated such that the mean vector has nonzero entries only for those indices in which the string is nonzero. The bias prior covariance is diagonal, and constructed in the same way as the mean. We will sometimes refer to this style of hypothesis set as ``considering all diagonal hypotheses.'' Parameter choices closer to the bottom left corner of the plot produce bias priors that are, in a formal sense, close to the null hypothesis.\footnote{In the binary case (two hypotheses), the plotted quantity is known as the ``Jensen-Shannon Divergence," and its square root satisfies the properties of a mathematical metric.} Parameter choices for which the mutual information is low indicate choices of hypotheses that are not reliably distinguishable from one another. Regions in which the mutual information is nearly maximal (4 bits in this example) indicate hypothesis sets for which the test can reliably decide which of the hypotheses is most likely. The corresponding posterior distribution over the hypotheses will usually be strongly concentrated at this hypothesis, i.e. it will be a low entropy state.


With only a few data points, Figure \ref{fig:mut_info} suggests that we may only be able to probe scenarios in which biases are frequently more than a few multiples of the error bar. In other words, it may only decisively point out visibly obvious statistical tensions. We remark that this is a rigorous, quantitative alternative to ``$\chi$-by-eye" that allows an analyst to tune and objectively assess statistical questions with confidence. For the HERA use case, there are hundreds of groups of 2-4 data points that would need to be assessed in this way, and so this test acts as a form of data reduction. We also point out that an easy avenue for combining more data and therefore increasing the information provided by the test is to combine wave modes in the power spectrum. Formalism for this procedure is contained in Appendix \ref{app:multi_post}.

In considering a set of priors for the alternate hypotheses, a useful choice for small data sets is one in which every combination of bias configurations is considered (as used in Figure \ref{fig:mut_info}). If zero-mean priors are used, this requires $2^N$ hypotheses, while nonzero-mean priors may need as many as $3^N$ hypotheses if all combinations of positive and negative biases for that mean are probed (assuming a single mean is used). If the analyst possesses genuine prior information that can eliminate some of these possibilities, then clearly some of these hypotheses need not be considered. Once a set of biased data has been identified, it may then be useful to introduce a second stage in the test, where the corresponding covariance matrix of the maximum a posteriori hypothesis is partitioned to identify common biases among the biased subset (Appendix \ref{sec:app_part}). For larger data sets where this exponential growth in hypotheses is impractical, one can imagine repeatedly bisecting the data and running the test with a small hypothesis set until a confident positive result for a bias configuration is located.\footnote{One notion of ``confidence" is that the maximum a posteriori hypothesis ``dominates" the other hypotheses with a substantial odds value, say 10:1 for all hypotheses. This naturally leads to using entropy as a measure of confidence.} For example, one could use only two hypotheses: one with no biases, and one where all epochs are biased independently. Since data where only some fraction of its points are biased can still be more consistent with this alternate hypothesis than the null one, this can be a useful way to reduce a data set with a coarser hypothesis test before applying a more targeted one.

The bias priors are meant to reflect hypothetical scenarios that the analyst is interested in knowing. Furthermore, since only a finite number of hypotheses is considered, the maximum a posteriori decision rule can only choose the hypothesis that is closest to modeling the data within the context of the considered hypotheses. An important implication of this concept is that even if a prior corresponding to some scenario seems to grossly misrepresent the content of the data, it may yet be a better match than, say, the null hypothesis. For example, having an excessively broad prior, e.g. greater than 70 times the size of the error bar, but zero mean, does not prohibit the decision rule from classifying a 3$\sigma$ outlier as a significant bias. This means that in order to assess fairly general propositions, such as ``the data contain at least one bias,'' one need not have an exhaustively precise model of how that bias should look, so long as the alternate hypothesis bias priors capture the meaning of ``significance" intended by the analyst (and the analyst is cautioned not to consider hypotheses indistinguishable from the null as being significant). In other words, the marginalization procedure (over the bias priors) endows the test with a degree of flexibility that allows the analyst to sculpt the test to be most sensitive to whichever scenarios they are most concerned about. However, the corresponding answer from the test is limited in scope, in that it can only judge whether some hypothesis is \textit{more likely} than another in consideration, e.g. ``it is more likely that there is at least one bias than that there are none, when these are the types of bias configurations in consideration.'' 


\subsubsection{Choosing the prior probability for the hypotheses}

The final choice of prior is that regarding each hypothesis, $P(\HH)$, i.e. the prior probability that each of the (mutually exclusive) hypotheses is true. This is a primarily subjective prior unless one has a model for how often anomalous, unexpected biases should occur in the data. There are many conceivable choices depending on the preferred line of questioning. For instance, one might adopt a flat prior over all bias configurations in order to show no preference to any one configuration. However, this will give preference to conclusions regarding certain overall numbers of biases. Mathematically, there is only one way for all $N$ data points to be simultaneously biased (though perhaps at a different level in each datum), but there are $\binom{N}{k}$ ways for there to exist $k$ biases within $N$ data points. This means that with a flat prior over the $2^N$ diagonal hypotheses, if one asks the question ``how many data are biased?" rather than ``which combination of biases is most like the data?" then they are a priori more likely to find an answer closer to $N/2$ (where the combinatorial factor is highest) than something closer to $N$ or 0. A similar statement holds for the various correlated partitions of any one of these bias macrostates (Appendix \ref{sec:app_part}). In what follows, we choose flat priors at every stage. For the HERA problem, this expresses our assumption that each epoch is just as likely to be biased as each other epoch, which we reason from the fact that data from each epoch was processed identically (exposed to the same selection cuts and systematic mitigation methods).

In subsequent sections, we apply these results to an array of examples of ranging realism. 

\begin{figure*}
    \centering
    \includegraphics[scale=0.6]{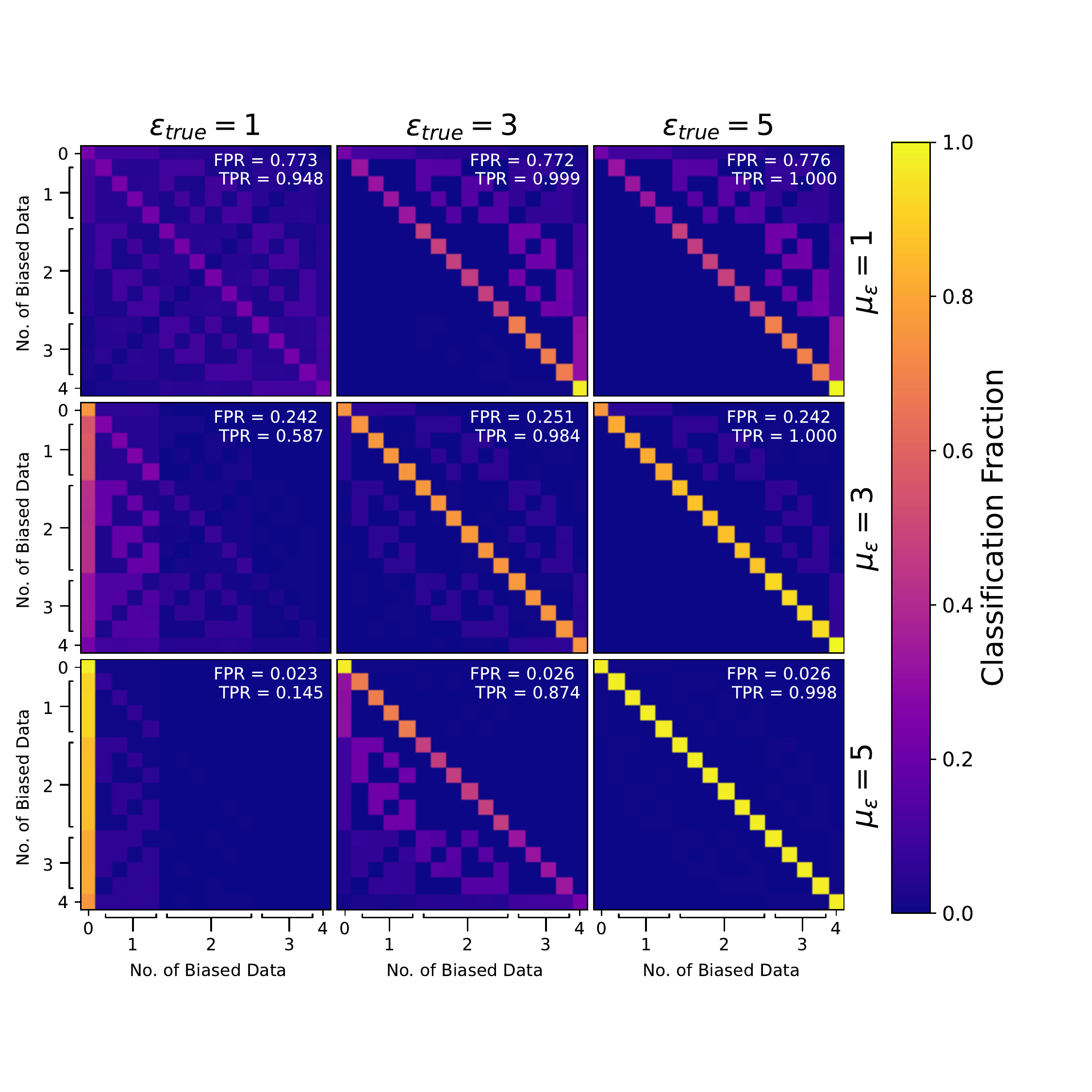}
    \caption{Classification matrices for the $N=4$ system investigating the 16 basic bias combinations. The bias strength in the data is given by the column of the panel (labels at the top of each column), and the assumed bias strength is given by the row (labels at the right of each row), each taking the values 1, 3, and 5 in units of the noise standard deviation. Within each panel is a $16\times16$ matrix, where the column index indicates the particular bias configuration that was chosen by the MAP rule, and the row index indicates the actual simulated bias configuration. The color indicates the fraction of samples simulated according to a particular state (row index) that were classified as the state indicated by the column index. In other words, diagonal entries represent proper identifications, and off-diagonal entries represent classification errors of some form. The annotations beside the axes tell the number of biases in the indicated state. In the upper right corner of each panel is listed the ``false positive rate" (fraction of unbiased data classified as somehow biased) and the ``true positive rate" (fraction of biased data classified as containing at least one bias, even if the biased states are incorrectly identified). When the parameters describing the alternate hypotheses do not lie in the region of high mutual information, there is an extremely large false positive rate. While using alternate hypotheses that are highly distinguishable generically decreases the true positive rate, the ratio of true positives to false positives increases, indicating that one can be more confident in positive classifications with this choice of hypothesis set.}
    \label{fig:toy_class}
\end{figure*}

\section{Results with Toy Data}
\label{sec:toy}

In this section, we apply the formalism to some basic simulations to build intuition. We then apply different decision rules to the posteriors and subsequently examine their performance as classifiers. In these first examples, we examine the case when only a few data are combined, since this is the most analogous to the HERA example. 

\subsection{Formulating a Decision Rule}

Once the posterior probabilities of each hypothesis are calculated, one may then be interested in formulating a decision about which hypothesis to assume. A decision problem is typically formulated by determining an objective function that is extremized according to the goals of the analyst. We define a loss function, $L(\HH_i, \HH_j)$, and minimize the expected loss over the posterior. Formally,
\begin{equation}
    \langle L(\HH_i) \rangle = \sum_{j=0}^{Q - 1}L(\mathcal{H}_i, \mathcal{H}_j)P(\HH_j | \bold{d} , \bold{C}_0).
    \label{eq:exp_loss}
\end{equation}
The quantity $L(\HH_i, \HH_j)$ describes the specific loss associated with behaving as if $\HH_i$ is true when in fact $\HH_j$ is true. $L(\HH_i)$ is then the expected loss of behaving as if $\HH_i$ is true. 

Defining a loss function is sometimes arbitrary, although an objective basis for one can also exist. For instance, in the HERA problem, one might ascribe a loss function based on how much wasted reanalysis occurs when an epoch is misidentified as containing a bias, assuming one goes back to check on the data through some independent analysis when a bias is identified. The usual loss function chosen in a wide variety of analyses throughout the general scientific literature is the sum of squared errors. To apply this to our problem we would need to formulate some sort of ``squared distance" between our various hypotheses. There are various candidates from the divergences of information theory, e.g. the Kullback-Liebler divergence or the Jensen-Shannon divergence between any two hypotheses. The latter is equivalent to the mutual information in a binary test between the two hypotheses in question. If we choose a set of hypotheses that are all mutually distinguishable as advised in \S\ref{sec:parameter_choice}, then the squared distance (measured in bits in this case) will be approximately 1 bit between any two hypotheses. In other words, there will be equal loss for any misclassification produced by the algorithm, and 0 loss for a proper classification (identifying exactly which epochs are biased in the HERA problem). Using such a loss function, we would obtain
\begin{equation}
    \langle L(\HH_i) \rangle = 1 - P(\HH_i | \bold{d}, \bold{C}_0)
    \label{eq:MAP_loss}
\end{equation}
which is minimized for the hypothesis whose posterior probability is maximal. In other words, this just reproduces the maximum a posteriori (MAP) decision rule.\footnote{Even if the hypotheses are not all highly distinguishable, some hypothesis sets result in roughly equal Jensen-Shannon divergences anyway, which will also produce the MAP rule.}

A loss function of particular relevance to the HERA problem might be to penalize false negatives more severely compared to other misclassifications. While identifying a bias where there are none produces concern for the quality of the power spectrum limit, a false negative fails to prevent a lower-quality limit. On the other hand, the analysts may also worry about producing selection bias, in which case a loss function that minimizes false positives would be relevant. In general, there is a tradeoff between false negatives and false positives, as we show in exploration below.

\subsection{Classification Performance}

We explore the classification performance of the jackknife test by simulating large ensembles of biased data and examining various metrics. First we use the loss function of Equation \ref{eq:MAP_loss} (MAP decision rule) and count how many instances of each bias configuration are mapped into each bias hypothesis. We also examine this ensemble in terms of entropy. We then inspect the true and false positive rates of a decision rule where false negatives are strongly penalized. 

In Figure \ref{fig:toy_class}, we show an array of ``confusion matrices." We generate an ensemble of Gaussian random vectors of length $N=4$ under each of the 16 possible bias configurations, with a standard deviation of 1 in each dimension and no correlations. We choose a true bias of (exactly) 1, 3, and 5, i.e. $\varepsilon_{\text{true}} \in \{1, 3, 5\}$.
We choose a set of alternate hypotheses with $\mu_\varepsilon$ equal to 1, 3, and 5 as well, and $\bold{C}_{\eps}$ nearly 0. Then, for each simulated data set and choice of hypotheses, we count how many data are most likely to come from each of the various hypotheses. From this, we can also calculate the false positive rate, which is the fraction of unbiased sample vectors that are classified as containing at least one biased component, as well as the true positive rate, which we define as the fraction of vectors with at least one component identified as biased (even if the truly biased components are misidentified). All off-diagonal entries in the matrix represent misclassifications, and the diagonal entries show proper classifications. 

\begin{figure}
    \centering
    \includegraphics[width=\linewidth]{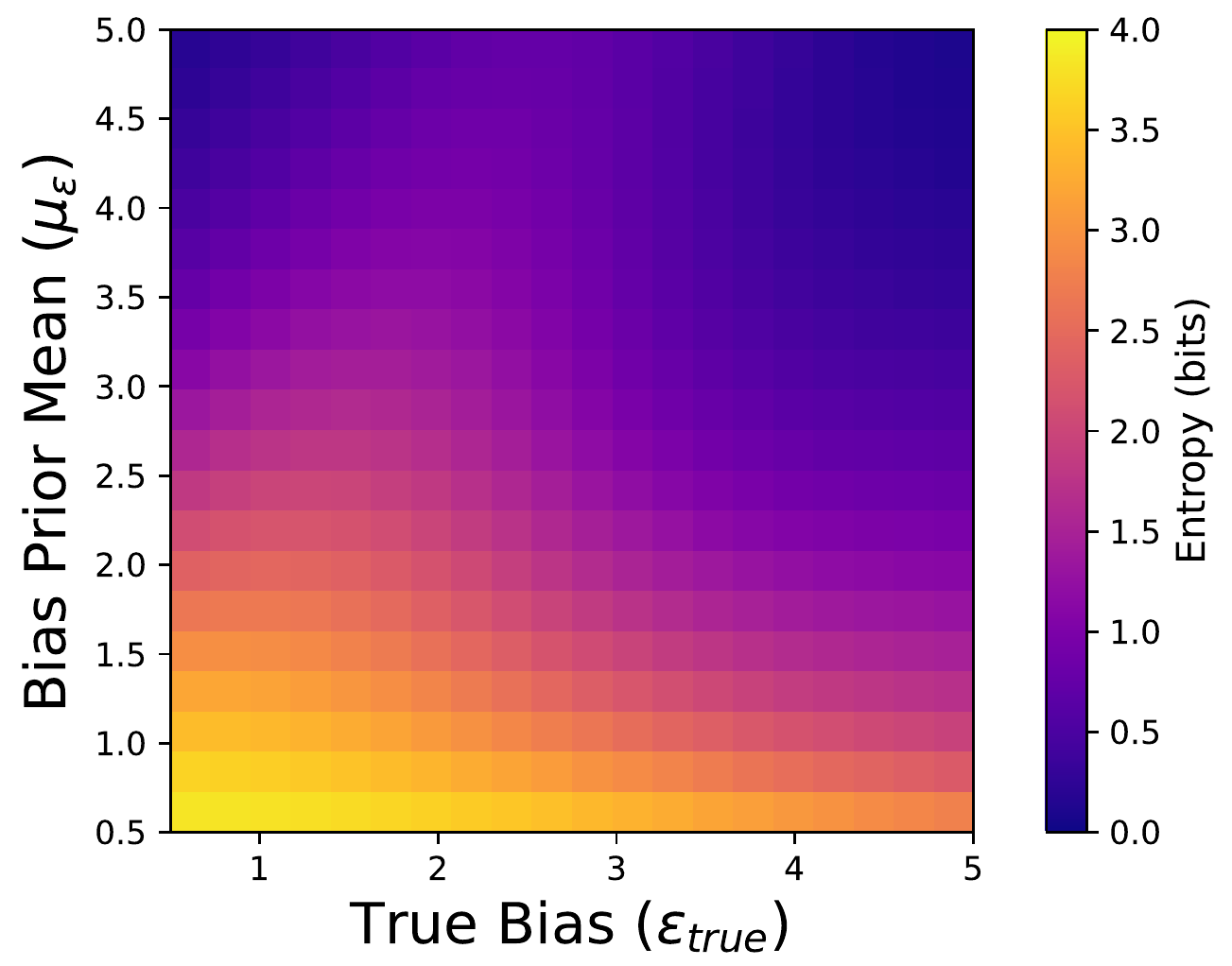}
    \caption{Entropy of the posterior distribution under different choices of alternative hypotheses and true biases. Regions of high entropy towards the lower left reflect uncertain conclusions, while regions of lower entropy express more certainty (generally fewer competing hypotheses).}
    \label{fig:avg_entropy}
\end{figure}

Unsurprisingly, when biases are strong (right-most column), the true positive rate is extremely high. However, if the alternate hypotheses are not sufficiently distinct from the null hypothesis (upper row), the MAP rule suffers from frequent misclassification within the biased states. The problem is more prominent for configurations with fewer biases. In addition, the false positive rate is extremely large. The false positive rate is, in general, not a function of the true biases, since the null hypothesis is identical regardless of $\varepsilon_\text{true}$. The extra aggression exemplified in the top row is also manifest in the pattern of misclassified true positives. When such misclassifications occur, they generally conclude that there are more biases present than in reality. In other words, for the top row, the most common misclassification is that states with one bias are classified as having two biases, and two bias states are misclassified as having three, etc. Usually the correct bias components are identified, but an extra bias is proposed. This is caused by an upward statistical fluctuation in the unbiased state appearing as a bias, which is a significantly less likely scenario when the assumed bias level is large compared to the noise. For the bottom row of matrices, the misclassification pattern is the opposite, except for when biases are extremely weak compared to the alternate hypotheses. When biased data are misclassified, they are often classified as containing one less bias then they truly possess. However, once the biases are sufficiently weak, the most common misclassification is a false negative (bottom left matrix). Such confusion matrices offer a way of choosing hyperparameters for the alternative hypotheses in terms of one's appetite for various types of failures, rather than from the theoretical considerations presented in \S\ref{sec:parameter_choice}. However in general we see that these two methods correspond with one another in a logical manner.

\begin{figure*}
    \centering
    \includegraphics[scale=0.52]{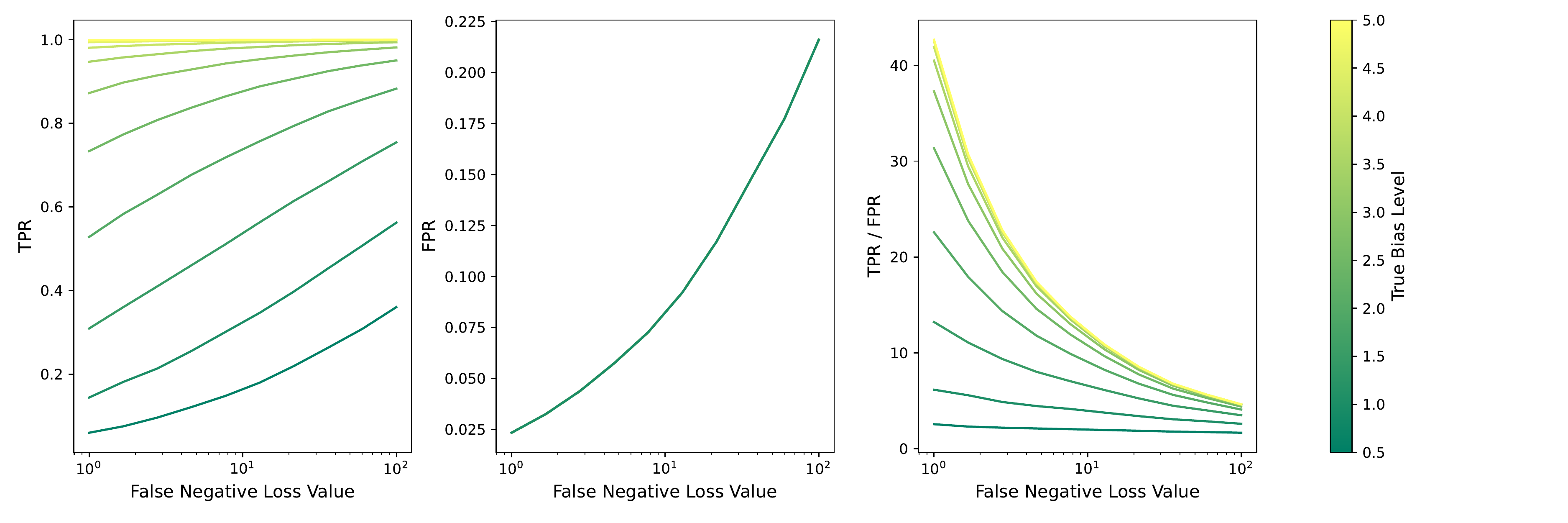}
    \caption{True positive rates, false positive rates, and their ratio when using a bias prior mean of 5 times the error bar, as a function of the strength of loss for false negatives (Equation \ref{eq:L_FNR}). For larger loss values, we can obtain strong gains in true positive rates for smaller biases in exchange for a larger false positive rate. The odds of true to false positives decrease monotonically as the loss strength goes up, but reasonable odds can be maintained for relatively weak biases at an appreciable increase in true positive rate.}
    \label{fig:FPR_TPR}
\end{figure*}

In Figure \ref{fig:avg_entropy}, we use entropy as a measure of the ``confidence" of the posterior distribution. The basic idea is that if the posterior is strongly concentrated on a small number of states, the sample entropy for that data draw
\begin{equation}
    \hat{S}(\HH | \bold{d}^{(i)}) \equiv -\sum_{j=0}^{Q-1}P(\HH_j | \bold{d}^{(i)})\log_2\big(P(\HH_j | \bold{d}^{(i)})\big)
\end{equation}
will be low. If we average over all $\bold{d}^{(i)}$ in the ensemble, then we have the sample entropy of the data set under that particular data generating distribution,
\begin{equation}
    \hat{S}(\HH | \bold{d}) \equiv -\frac{1}{D}\sum_{i=1}^D\hat{S}(\HH | \bold{d}^{(i)}),
    \label{eq:samp_ent_2}
\end{equation}
$D$ being the number of samples drawn. This acts as a measure of how certain the algorithm is of its classifications on average. We repeat a similar simulation process as for Figure \ref{fig:toy_class}, but on a more finely spaced grid of $\mu_\varepsilon$ and $\varepsilon_\text{true}$. Additionally, we mix all bias configurations at a given $\varepsilon_\text{true}$ in equal proportions. We generally observe lower entropies for higher $\mu_\varepsilon$ and $\varepsilon_\text{true}$. For low $\varepsilon_\text{true}$, and high $\mu_\varepsilon$, these low entropy states are ``confident misclassifications." This is reasonable behavior, since weak biases should be difficult to distinguish from noise when one is expecting strong biases. For example, if one is specifically expecting $5\sigma$ biases and observes a datum at $1\sigma$, it would be unreasonable to be in doubt about which hypothesis is most likely. In a complementary vein, if one expects weak biases and weak biases are often present, one should expect to be highly uncertain about the state of the system. This corresponds to the region of high entropy in the lower left of the plot.

So far, we have not made much use of the loss function formalism. A loss function is highly contextual. A particularly important type of misclassification in the HERA problem is a false negative (i.e. concluding no biases when at least one is present). If a bias can be identified before coherently averaging the data to form a final power spectrum, then in principle the data that led to the subspectra on which the test is being run can be reanalyzed and potentially cleared of systematics. Conversely, failing to identify a bias can lead to a systematically dominated power spectrum upper limit that could have otherwise been avoided. One may use the loss function in order to more severely penalize such misclassifications and therefore avoid them more often in exchange for more false positives. Alternatively, one might be concerned with selection bias effects, such as in the HERA problem where we deploy the jackknife test on a measurement that is extremely close to the final scientific product. In that case, one could construct a loss function where false positives are punished more severely, so that one does not inadvertently label unbiased data as biased.

To illustrate this with math, we can think of Equation \ref{eq:exp_loss} as a loss matrix $\bold{L}$ acting on a posterior probability vector $\bold{p}$, where the $ij$th entry of $\bold{L}$ is equal to $L(\HH_i, \HH_j)$ and the $j$th component of $\bold{p}$ is the posterior probability of the $j$th hypothesis. To achieve the MAP rule, one would use a loss function given by
\begin{equation}
    \bold{L}_\text{MAP} = \mathds{1}\mathds{1}^T - \bold{I}_Q =  \begin{pmatrix}
    0 & 1 & 1 & ... & 1 \\
    1 & 0 & 1 & ... & 1 \\
    1 & 1 & 0 & ... & 1 \\
    \vdots &&& \ddots & 
    \end{pmatrix},
\end{equation}
where $\bold{I}_Q$ is the $Q\times Q$ identity matrix. This loss matrix is a symmetric matrix with 0s down the diagonals and 1 for all other entries, i.e. 0 loss for proper classifications and a unit loss for any misclassification regardless of its type. This reproduces Equation \ref{eq:MAP_loss} exactly. To produce a lower false negative rate at the expense of more false positives, one changes the top row of this matrix to some false negative loss value, $L > 1$, except for the diagonal entry. In math,
\begin{equation}
   \bold{L}_\text{low FNR} = \begin{pmatrix}
    0 & L & L & ... & L \\
    1 & 0 & 1 & ... & 1 \\
    1 & 1 & 0 & ... & 1 \\
    \vdots &&& \ddots & 
    \end{pmatrix}.
    \label{eq:L_FNR}
\end{equation}
Similarly, to achieve a lower false positive rate, at the expense for more false negatives, one replaces the first column with some loss value, $L > 1$. Explicitly,
\begin{equation}
    \bold{L}_\text{low FPR} = \begin{pmatrix}
    0 & 1 & 1 & ... & 1 \\
    L & 0 & 1 & ... & 1 \\
    L & 1 & 0 & ... & 1 \\
    \vdots &&& \ddots & 
    \end{pmatrix}.
\end{equation}

In Figure \ref{fig:FPR_TPR}, we explore the effect of modulating a false negative penalty while holding all other misclassifications at a constant loss value. In other words, we vary $L$ in equation \ref{eq:L_FNR}. When $L = 1$, this is the same as the MAP rule. Using the same simulation suite as Figure \ref{fig:avg_entropy}, we analyze the true and false positive rate as we adjust the strength of the loss for false negatives relative to that of other misclassifications. We only use alternate hypotheses with $\mu_\varepsilon=5\sigma$. The left-hand panel shows the true positive rate as a function of loss value for the chosen biases. We see that for weak biases we can appreciably increase the true positive rate. The middle panel shows the false positive rate. Combining this with the left-hand panel, one may tune the false positive rate and then estimate the sensitivity to various biases. For instance, using a loss strength of 10 for false negatives increases the true positive rate for $1\sigma$ biases by about 50\% of its MAP value (loss strength equal to 1) while keeping the false positive rate at less than 10\%. The right-hand panel shows the ratio of true to false positive rate. If for some reason the prevalence of each bias configuration is known (or if one is willing to use prior values), this quickly translates to an odds of a positive result being true versus false. This can also be used to tune the loss function by setting an odds threshold for a given strength of bias. In general, we see that this ratio is monotonically decreasing as a function of loss strength, indicating that the false positive rate generally increases quicker than the true positive rate.

Using a large array of simulated data has allowed us to thoroughly explore the statistical properties underlying this test. Specifically, we have illustrated a number of tools one can use in order to tune the test to the question at hand. We have primarily focused on the form of the jackknife test in which one is interested in all $2^N$ bias configurations without asking about interbias similarities ($\bold{C}_{\eps}$ diagonal). In principle, the manner of investigation presented in this section may be used for any hypothesis set, e.g. those formalized in Appendix \ref{sec:app_part}, or a more specific hypothesis set defined by the analyst's problem.  We first turn to simulations from the HERA validation pipeline in order to investigate the performance of the test in a controlled yet realistic setting, and then examine the results of the test on the measured HERA bandpowers.

\begin{figure}
    \centering
    \includegraphics[width=\linewidth]{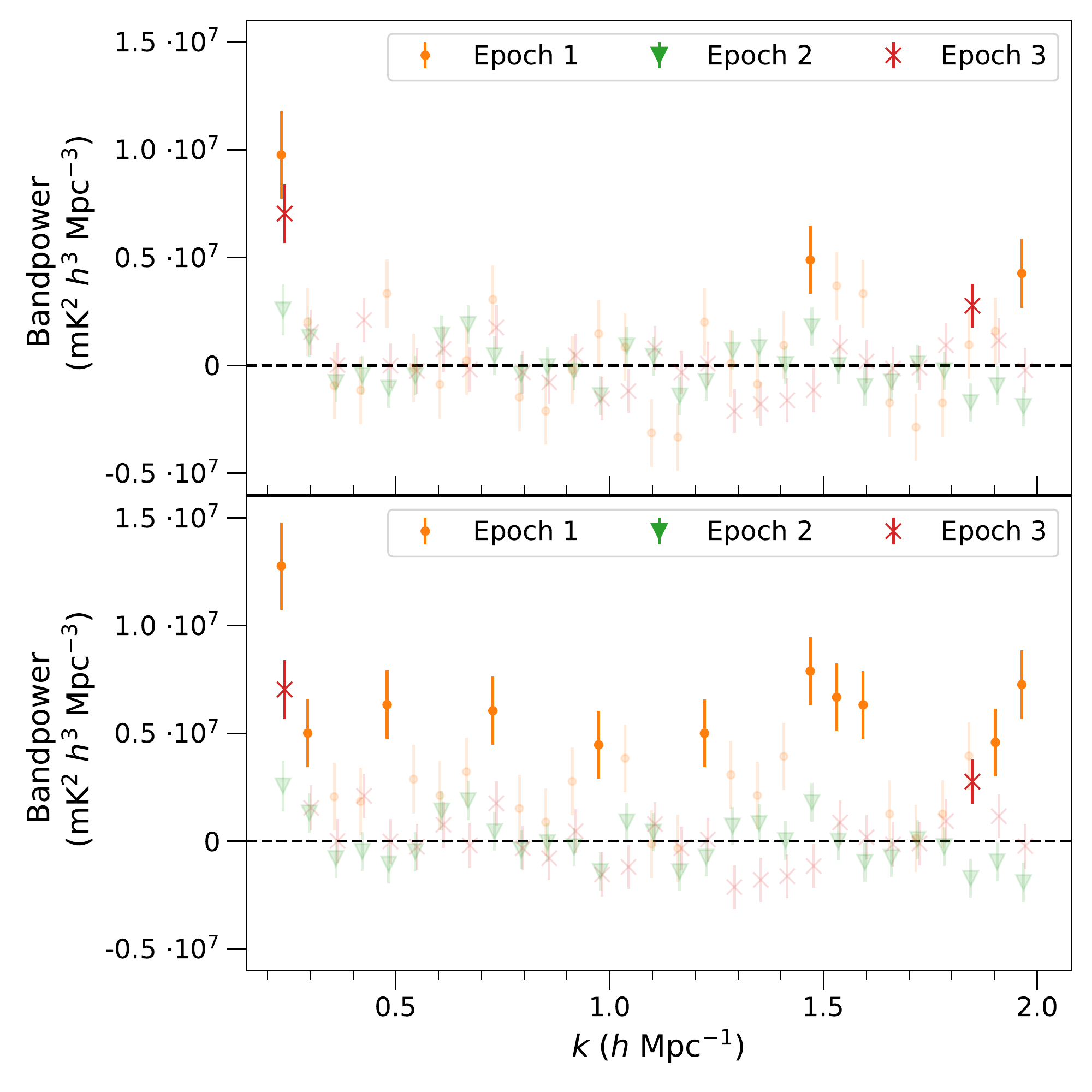}
    \caption{Top: Bandpowers directly from the HERA validation pipeline, simulating foregrounds, noise, systematics, and systematic subtraction (no EoR signal), for band 1, field E (top). Bottom: Same bandpowers, except we artificially add a bias to the epoch 1 measurements of $3\cdot10^6$ mK$^2$ $h^3 $ Mpc$^{-3}$ (approximately two times the size of the error bar for most modes). Opaque data have been identified as biased according to the MAP decision rule (see \S\ref{sec:valid} for test details).}
    \label{fig:valid_bp}
\end{figure}

\begin{figure}
    \centering
    \includegraphics[width=\linewidth]{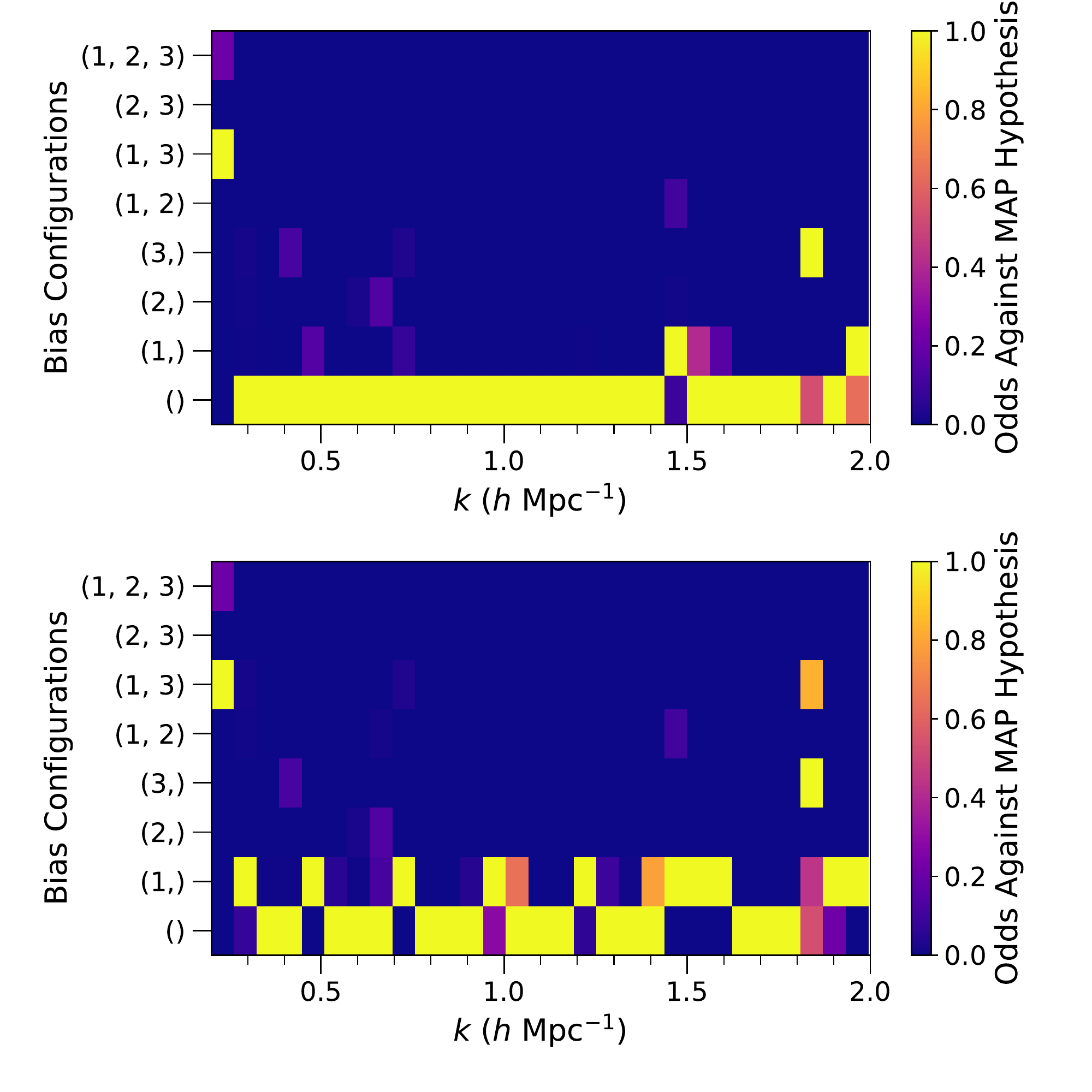}
    \caption{Top: The odds against the MAP hypothesis based on the posterior distribution for each mode when considering the raw validation bandpowers (top panel of Figure \ref{fig:valid_bp}). As expected, the raw validation results are largely consistent with zero-mean noise (the plot is concentrated at the empty bias configuration with relatively few competing hypotheses). Bottom: When a bias is artificially added to epoch 1, we see that the hypothesis that ``epoch 1 is biased" is highly favored compared to other hypotheses for about half the modes.}
    \label{fig:valid_stoplight}
\end{figure}

\section{Results for HERA Simulations and Data}
\label{sec:HERA_data}

In this section, we apply the jackknife test to the simulated and measured HERA bandpowers from their first observing season, as constructed and analyzed in \citet{HERA2022b}. The data for the observing season, spanning 94 nights after initial data quality cuts, was split into four epochs of roughly equal size. The data were then analyzed identically, and power spectra were formed on a per-epoch basis. Accompanying the HERA data analysis pipeline is a comprehensive validation pipeline designed to verify the efficacy of the data analysis techniques using thorough and realistic simulations of the experiment \citep{Aguirre2022}. This includes simulation of foregrounds, fiducial EoR signals, thermal noise, calibration, known systematic effects, systematic mitigation techniques, and so on. For \citet{HERA2022b}, all 94 nights passing initial data selection were simulated and separated into epochs, mimicking the exact data analysis procedure. This allows us to apply the jackknife test to both the simulated bandpowers as well as the measured bandpowers. 

We first show the results of the jackknife test on validation simulations where foregrounds are simulated, but no EoR signal is present. This effectively produces a data set where the null hypothesis is satisfied almost everywhere. We then modify the outputs of these simulations by adding a constant bias to one of the epochs on all modes, which is an effect that could arise, for example, due to the presence of ultra-faint RFI \citep{Wilensky2020}. Note we do not actually use the validation pipeline to simulate the effect of RFI (or any other systematic that might add a constant bias); we are just artificially introducing the bias directly to the end-stage bandpowers. Finally, we show the results of the jackknife test on the actual measured bandpowers, demonstrating a majority null result, thereby justifying (post hoc) the decision to coherently average the epochs together to form one final power spectrum. 

\subsection{Results with HERA Validation Simulations}
\label{sec:valid}

In Figures \ref{fig:valid_bp} and \ref{fig:valid_stoplight}, we show the results of the jackknife test for a single band and field of power spectra made using the HERA validation pipeline. These simulations included foregrounds, noise, instrumental/analysis systematics, and systematic mitigation, but no EoR signal. The jackknife settings are contained in the bulleted list below.
\begin{itemize}
    \item \textbf{Hypothesis Set}: All Diagonal
    \item \textbf{Bias Mean/Width}: $6\sigma_i \pm \sigma_i$
    \item \textbf{Null Mean/Width}: $0 \pm 0$
    \item \textbf{Decision Rule}: MAP
\end{itemize}
We test all diagonal hypotheses (i.e. all epoch combinations, but without asking whether some are degenerate as in Appendix \ref{sec:app_part}), using a prior mean and width that are sized based on the per-epoch error bar. We use the MAP decision rule for determining biased epochs, which is equivalent in the preceding section's formalism to using a constant loss function over the range of possible misclassifications. Figure \ref{fig:valid_bp} shows the bandpowers with (bottom) and without (top) the addition of an artificial bias to epoch 1 that is roughly two times the size of the error bar for most modes. Figure \ref{fig:valid_stoplight} shows the posterior probability of each bias configuration divided by the posterior probability of the MAP hypothesis, i.e. the odds against the MAP hypothesis. Such a plot allows us to see how competitive various hypotheses are with the MAP hypothesis. This shows us that realistic simulations intended to be somewhat consistent with the null hypothesis at sufficiently high $k$ do appear to demonstrate this consistency. Since small residual systematics still exist despite mitigation, we do not expect perfect consistency with the null hypothesis. When we artificially add a bias to epoch 1, the MAP solution shifts towards identifying epoch 1 as biased for about half the modes, as expected. 



\begin{figure*}
    \centering
    \includegraphics[scale=0.225]{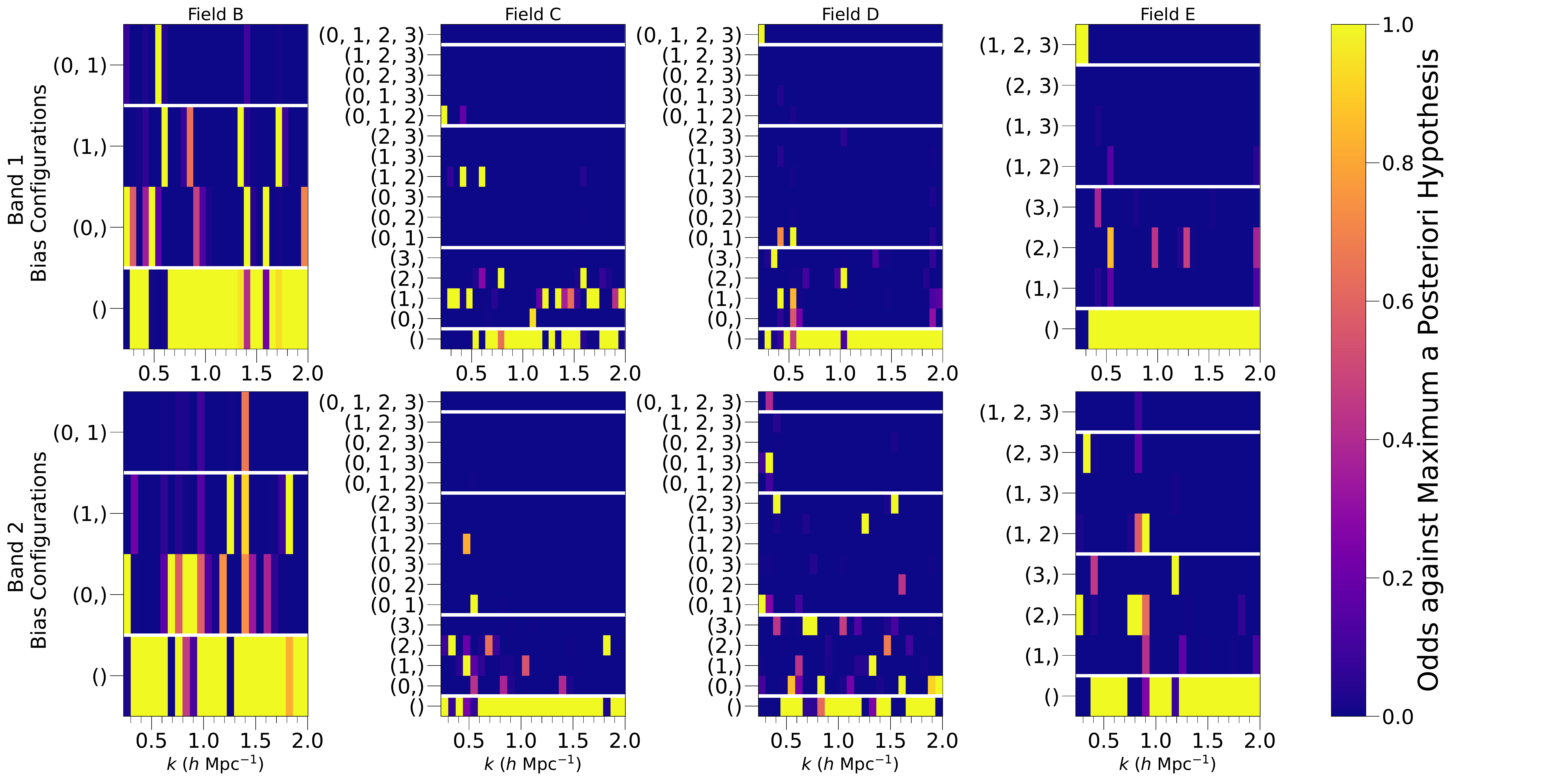}
    \caption{Odds against the maximum a posteriori hypothesis for (positive) bias configurations (vertical axes -- contents of the tuple indicate which epochs are biased under the hypothesis) in the measured bandpowers for each observed frequency band (panel rows), celestial field (panel columns), and spherical wave number (horizontal axes). The horizontal white lines delineate the borders between sections of the plot where certain numbers of epochs are biased. We observe that the null hypothesis is most likely in the majority of instances, suggesting that the data are unlikely to contain significant biases in the bandpowers for most bands, fields, and modes. Interestingly, epoch 1 is often identified as a biased bandpower in band 1, field C with few or no competing hypotheses on most modes. This can be visually confirmed in Figure \ref{fig:meas_bp}, where epoch 1 demonstrates a consistent positive bias across most $k$-modes.}
    \label{fig:stoplight}
\end{figure*}

With these settings, we expect a false positive rate of about 1.5\%, and a true positive rate for biases of $2\sigma$ of 49\%. When we artificially add a $2\sigma$ bias, the data are roughly consistent with this true positive rate. Examining validation outputs from all bands and fields, and accounting for differences in number of hypotheses between tests, we find that there are more data identified as biased than expected from the false positive rate, which is between 1 and 2\% depending on the field (higher rates for more epochs). About 10\% of the validation bandpowers over all bands and fields are identified as biased. Slightly less than half of the identified bandpowers are in the lowest $k$-mode. While efforts have been made to exclude contributions from the foreground wedge to these modes, it is possible that the cuts that provide these exclusions \citep{HERA2022b} may be imperfect, and that some residual foreground signal exists in these bins. The remaining identified data points appear somewhat uniformly scattered throughout the rest of the wave numbers. This excess above the false positive rate suggests a potential low-level violation of the null hypothesis, such as residual systematics that are incompletely subtracted. For example, biases at $1/2$ the error bar have a true positive rate of about 13\%. With the exception of these potential nuances, we observe that the jackknife test performs more or less within expectation and proceed to applying it on the actual measured bandpowers.

\subsection{Results with HERA Data}

Figure \ref{fig:stoplight} shows the posterior odds against the MAP hypothesis for each wave mode in the power spectra of the observed bands and fields in \citet{HERA2022b}. There are two bands: one extending over 117.19--133.11 MHz and another extending over 152.25--167.97 MHz. The fields span roughly 17 hours of right ascension, with two small gaps between fields A and B as well as B and C. Fields C, D, and E are contiguous. They are not all equally sized, both in terms of angular extent and total data volume contributing. Additionally, due to the large angular extent of the field collection, some epochs do not have measurements for some fields. This is reflected in the thermal uncertainties of the measured bandpowers. For this test, we use the same bias hypotheses as in the validation test, whose parameters are listed in \S\ref{sec:valid}. The null hypothesis, that the data are consistent with zero-mean noise, is the MAP solution in the majority of instances, suggesting that the data are unlikely to possess substantial positive biases\footnote{We also searched for negative biases, for which the algorithm reports a small handful of negative outliers among all the data.} for most $k$-modes in the bands and fields that were tested. Furthermore, there is usually only a small number of competing hypotheses compared to the size of the configuration space. 

\begin{figure*}
    \centering
    \includegraphics[scale=0.4]{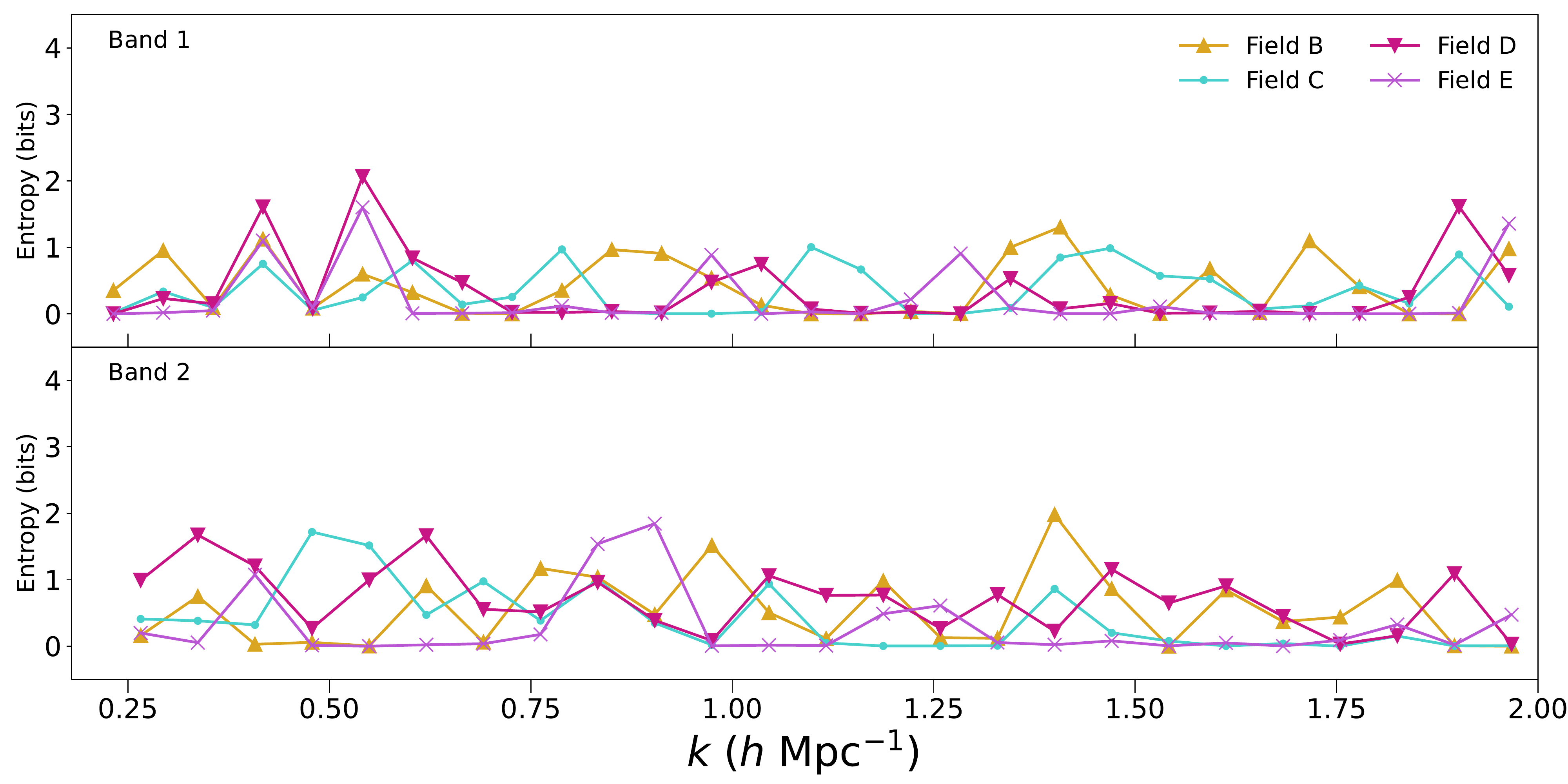}
    \caption{Entropy of the posterior for each band and field as a function of wave number. Points of low entropy correspond to posteriors with very few competing hypotheses, i.e. very certain inferences. Higher entropy states, such as band 2, field B, $k=1.40 h$ Mpc$^-1$, indicate uncertain inferences. As observable in Figure \ref{fig:meas_bp}, this particular mode has two positive-valued bandpowers slightly more than $2\sigma_i$ away from 0 -- dubious evidence of a potential bias. Ultimately the test reasons it is unbiased under the chosen test parameters.}
    \label{fig:ent_two_panel}
\end{figure*}

This is summarized neatly by the entropy of the posteriors, which we show in Figure \ref{fig:ent_two_panel}. In most instances, the entropy is less than 1 bit. Interpreting $2^H$ as the effective number of competing hypotheses, having less than 1 bit of entropy suggests that there is usually less than 1 hypothesis with comparable probability to the MAP solution, though this may be manifest as multiple other hypotheses each with low to moderate probability. The relatively low entropy of the posteriors is largely a consequence of choosing alternative hypotheses that are highly distinguishable by the data, cf. Figures \ref{fig:mut_info} and \ref{fig:avg_entropy}.  

\begin{figure*}
    \centering
    \includegraphics[scale=0.34]{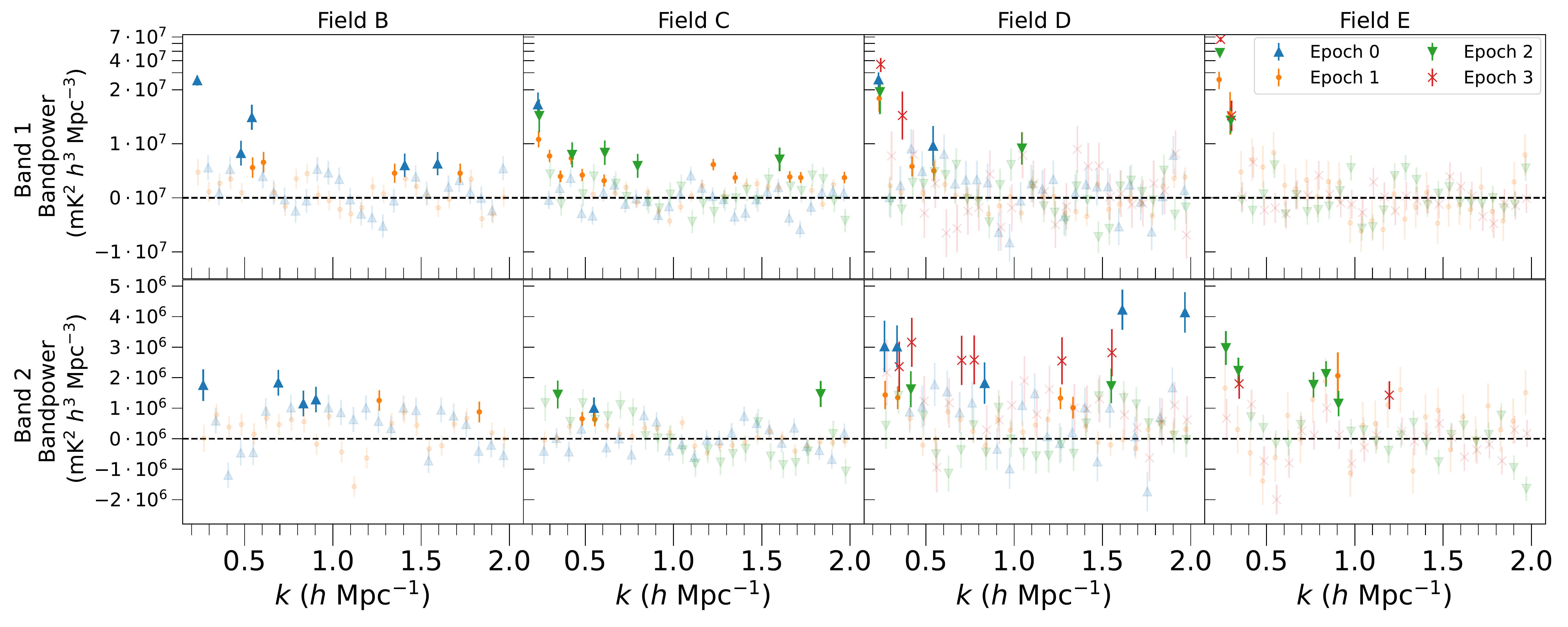}
    \caption{Measured bandpowers and $1\sigma$ error bars from \citet{HERA2022b}. In the top row, we have compressed the upper range of the plot using a logarithmic scale above $2\cdot 10^7$ in order to show some extreme outliers without obscuring the clarity at lower power. Data identified as positively biased according to the MAP decision rule are opaque, while data classified as unbiased are transparent. Points from different epochs are slightly staggered for visual clarity. Note that the two lowest limits belong to band 1, field D and band 2, field C, which are quite well-behaved under the jackknife test in general. The particular modes from which the lowest limits come ($k = 0.36$ $h$ $Mpc^{-1}$ and $k = 0.34$ $h$ $Mpc^{-1}$, respectively), do show strong evidence of biases in one epoch each. Note that in field C, we have removed epoch 3 from the plot for visual clarity, since it contained substantially less data in this particular field than other epochs, leading to large error bars that obscure the finer details in the plot.}
    \label{fig:meas_bp}
\end{figure*}

Figure \ref{fig:stoplight} also allows one to easily see whether a particular epoch is often problematic, such as epoch 1 in band 1, field C. Since the full posterior is available (normalized against the MAP), we can also read off trends in the competitive hypotheses. For instance in band 1, field B, the odds of a scenario where only epoch 1 is biased and the null hypothesis is close to 1 in a pair of high $k$-modes, and epoch 1 is highlighted for two other modes in the same panel. This suggests that if we were to combine information over $k$-modes by using the formalism in Appendix \ref{app:multi_post}, then we may observe fewer null results. 

In Figure \ref{fig:meas_bp}, we show the measured bandpowers for each band and field, along with $1\sigma$ error bars. Opaque data were those identified as biased according to the MAP solution, while transparent data are classified as unbiased. Comparing to Figure \ref{fig:stoplight}, we can visually confirm that epoch 1 consistently measures bandpowers that are several error widths above 0. Given that the false positive rate with these settings is roughly 6\%, we are confident that most of these identified biases are true positives. Most EoR models produce power spectrum signals that are orders of magnitude beneath the thermal noise on these modes, meaning that these biases are probably residual systematic effects. This plot highlights the ability of the jackknife test to quantitatively identify likely outliers in circumstances where visual identification would be difficult. We also remark that since the data are considered jointly and under several different models, the test is more surgical and gives different results than a simple $z$-score cut. Expanding on this point, the Bayesian formalism makes the conclusion more readily interpreted than a decision based solely on a $z$-score, which directly translates to a $p$-value under the null hypothesis. A $p$-value under one hypothesis is nonspecific to other models. This test provides more specific information about how the null hypothesis could be violated: it points the analyst in a particular direction when the null hypothesis is rejected. Moreover, it provides a specific criterion for when the null hypothesis should be rejected.

\begin{figure*}
    \centering
    \includegraphics[scale=0.38]{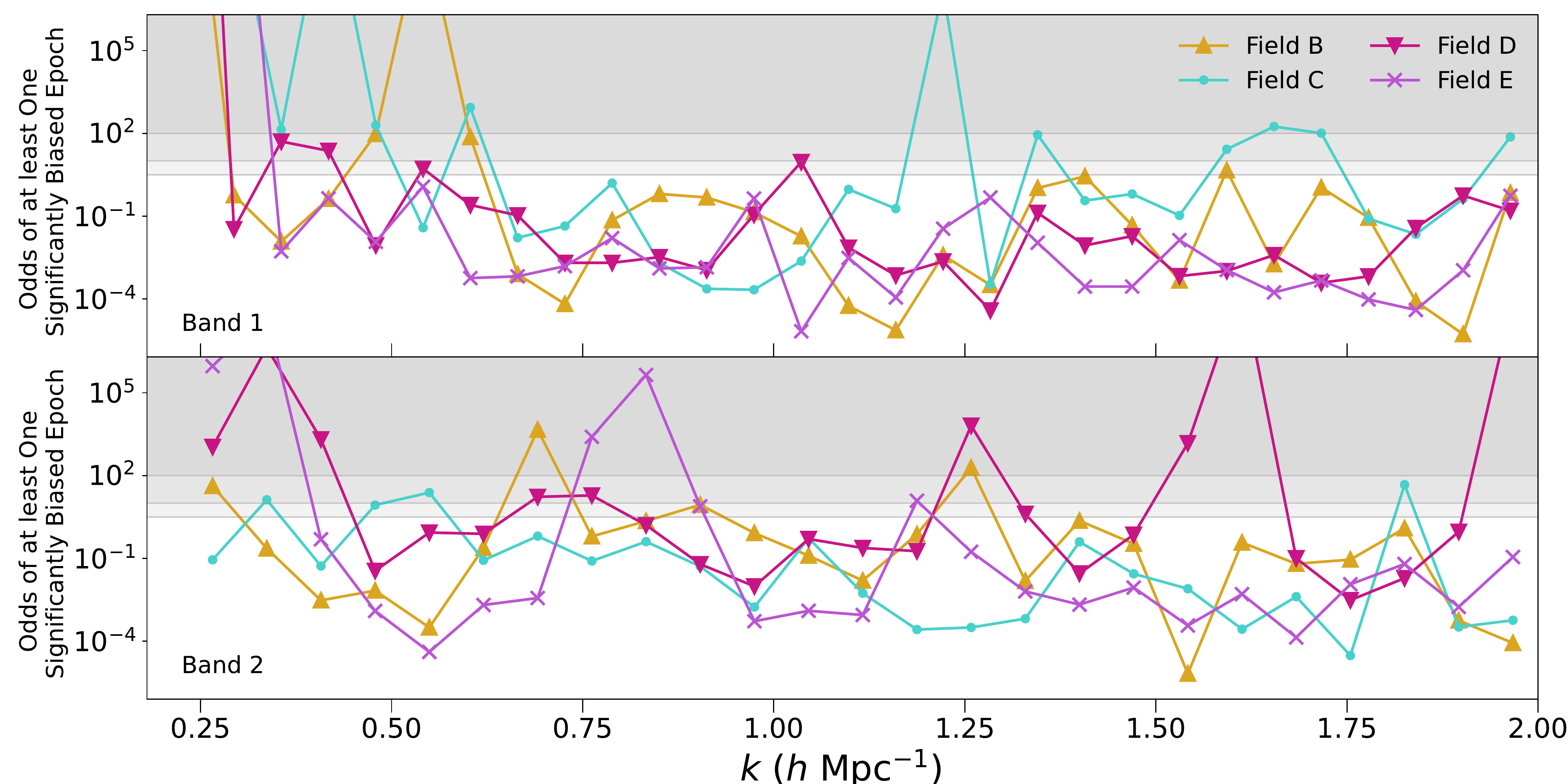}
    \caption{Odds of at least one significantly biased epoch in each band and field as a function of wave number (reproduced from \citet{HERA2022b}). This is calculated by summing the posterior over all bias configurations and dividing by the posterior probability of the null hypothesis. Regions on the plot are shaded according to the odds interpretation presented in \citet{Kass1995}. The majority of points show no decisive evidence of significant per-epoch biases (darkest shade, odds $> 10^2$), except at lower $k$-modes and with a few exceptions at higher $k$. A modest number of points lie in the region of strong evidence for bias (middle shade, odds between $10^1$ and $10^2$), and a handful of points lie in the small lightly shaded region between $10^{1/2}$ and $10^1$, which is a region of substantial but not strong evidence for bias. This majority null result in combination with the typically low entropies of the inferences essentially qualifies as a pass, except potentially for band 1, field C and band 2, field D.}
    \label{fig:odds_two_panel}
\end{figure*}

The decision to average the epochs together in each band and field in \citet{HERA2022b} was made prior to the application of this hypothesis test. Therefore, any justification that this test would lend to such a decision is post hoc, and serves strictly as validation of the decision. As with any decision, one may approach this in a variety of ways. For example, one might consider the power spectrum as a whole and make the decision based on a summary statistic that consistently responds for a large number of $k$-modes. Following this line of thinking, we construct the ``odds of at least one significantly biased epoch" per band, field, and mode by summing the posterior over all bias configurations and dividing by the posterior probability of the null hypothesis, displayed in Figure \ref{fig:odds_two_panel}. We might then decide a coherent average is unjustified if there is significant odds of a biased epoch on a large number of modes.\footnote{One could eliminate this mode counting step and generate posterior odds for the whole spectrum by applying the formalism in Appendix \ref{app:multi_post}.} Inspecting Figure \ref{fig:odds_two_panel}, we see that for most bands and fields, and modes, there is counterevidence (odds less than 1) or weak evidence of a significantly biased epoch. Exceptions include the lower-order $k$-modes, where we suspect some residual foreground contamination may be present, as well as band 1, field C and band 2, field D, which appear to have consistent evidence of bias on a large number of modes. In summary, we consider the result of this jackknife test as largely justifying the decision to coherently average the epochs together in \citet{HERA2022b}.

\section{Summary and Conclusions}
\label{sec:conc}

We presented the formalism for a jackknife test for identifying biases from a small collection of subsets of data. The Bayesian formalism is flexible in that different questions may be asked by posing different scenarios. These scenarios, or hypotheses, are described by prior probability distributions on the possible biases. The test essentially proceeds by calculating the evidence of each hypothesis. These evidences are marginal likelihoods of the hypotheses, and thus using Bayes' theorem, we can combine them into a posterior probability distribution over the hypotheses. A decision about whether a significant bias has been identified in a given scenario is made by specifying a loss function and minimizing its expected value over the posterior. 

We show the explicit calculation of the marginal likelihoods when Gaussian priors on the biases and underlying mean under the null hypothesis are used. This is not a necessary feature of the framework, but is appropriate in many cases and permits extremely fast computation the posteriors. This allows for rapid exploration of the performance qualities of the jackknife test, such as true and false positive rates. A key concept in our understanding of the test is Shannon entropy, which acts as a metric of certainty for the test. Related concepts such as mutual information allowed us to make a good theoretical guess for what types of bias hypotheses would provide informative jackknife tests. We then validated this theory by simulating the classification problem and evaluating confusion matrices for the simulated data. In principle, these exploratory techniques are available for whichever bias scenarios one may wish to probe, however hypotheses for which analytic marginalization fails will take longer in order to perform the relevant exploration.

A particular hypothesis set that we spend a great deal of time exploring is one in which any of $N$ data points may be biased by some unknown amount whose mean and (diagonal) covariance we specify. This produces $2^N$ hypotheses over which the posterior must be calculated. This is a fairly general hypothesis set that allows one to agnostically probe which data are most likely to be biased without specifying potential equivalencies between the biases. One may of course choose a subset of these hypotheses in order to sculpt a more specific question, or more generally, adjust the prior weights in order to favor some hypotheses over others. Furthermore, one may establish equivalence relations between the biases using the formalism that we present in Appendix~\ref{sec:app_part}. 

As a proof of concept, we applied the jackknife method to a HERA data set and accompanying validation simulations used for a power spectrum upper limit in \citet{HERA2022b}. The purpose of the test in this instance was to determine (post hoc) whether coherently combining subsets of the limit data was statistically justified. The basic idea is that if a particular subset of data consistently showed evidence of bias, then it would have been prudent to exclude that data from the coherent average since this would indicate that it had high levels of systematic contamination. We found that in most instances there was little evidence of consistent bias for any given epoch except in a minority of band/observing field combinations. We therefore conclude that the choice to coherently average the data together was indeed justified, except potentially at lower $k$-modes. In future applications to the HERA experiment, we aim to implement the multi-$k$-mode formalism in Appendix \ref{app:multi_post}, as well as apply the test in earlier stages of the pipeline such as the LST-averaged visibilities. We suspect that this jackknife test will be a valuable tool for assessing statistical tension in other cosmological data sets as well, for instance between different seasons or detector sets in CMB experiments, different dishes/baselines of 21cm intensity mapping experiments, and so on. Analytic and numerical versions of the test have been made available to the community in the \href{https://github.com/mwilensky768/chiborg}{\textsc{chiborg}} software.

\section*{Acknowledgements}

This material is based upon work supported by the National Science Foundation under grants \#1636646 and \#1836019 and institutional support from the HERA collaboration partners. 
This research is funded in part by the Gordon and Betty Moore Foundation through Grant GBMF5212 to the Massachusetts Institute of Technology.
HERA is hosted by the South African Radio Astronomy Observatory, which is a facility of the National Research Foundation, an agency of the Department of Science and Innovation. 
This result is part of a project that has received funding from the European Research Council (ERC) under the European Union's Horizon 2020 research and innovation programme (Grant agreement No. 948764; P.~Bull and M.J.~Wilensky).

P.~Bull and F.~Kennedy acknowledge support from STFC Grant ST/T000341/1.
J.S.~Dillon gratefully acknowledges the support of the NSF AAPF award \#1701536.
N.~Kern gratefully acknowledges support from the MIT Pappalardo fellowship.
P.~Kittiwisit and M.G.~Santos acknowledge support from the South African Radio Astronomy Observatory (SARAO; \url{www.sarao.ac.za}) and the National Research Foundation (Grant No.\ 84156).
Y.~Qin would like to acknowledge the support from the High Performance Computing centers of the Scuola Normale Superiore (Italy), the Council for Scientific and Industrial
Research (South Africa), the OzSTAR national facility (Australia) and XSEDE (U.S.) for computational resources.
J.~Mirocha acknowledges computational resources and support on the supercomputer Cedar at Simon Fraser University, which is managed by Compute Canada and funded by the Canada Foundation for Innovation (CFI).
Parts of this research were supported by the Australian Research Council Centre of Excellence for All Sky Astrophysics in 3 Dimensions (ASTRO 3D), through project number CE170100013.
S.~Heimersheim acknowledges support from STFC studentship 2277533 in project  ST/T505985/1.
J.B.~Mu\~noz is supported by a Clay Fellowship at the Smithsonian Astrophysical Observatory.
G.~Bernardi acknowledges funding from the INAF PRIN-SKA 2017 project 1.05.01.88.04 (FORECaST), support from the Ministero degli Affari Esteri della Cooperazione Internazionale -- Direzione Generale per la Promozione del Sistema Paese Progetto di Grande Rilevanza ZA18GR02 and the National Research Foundation of South Africa (Grant Number 113121) as part of the ISARP RADIOSKY2020 Joint Research Scheme, from the Royal Society and the Newton Fund under grant NA150184 and from the National Research Foundation of South Africa (grant No.\ 103424).
E.~de Lera Acedo acknowledges the funding support of the UKRI Science and Technology Facilities Council SKA grant.
A.~Liu acknowledges support from the New Frontiers in Research Fund Exploration grant program, the Canadian Institute for Advanced Research (CIFAR) Azrieli Global Scholars program, a Natural Sciences and Engineering Research Council of Canada (NSERC) Discovery Grant and a Discovery Launch Supplement, the Sloan Research Fellowship, and the William Dawson Scholarship at McGill.

\section*{Data availability}

The simulations used in \S\S\ref{sec:math}-\ref{sec:toy} can be reconstructed using example notebooks as part of the publicly-available \textsc{hera\_stats} repository.\footnote{\url{https://github.com/HERA-Team/hera_stats}} The validation outputs are available in the \textsc{h1c\_idr3\_validation\_notebooks} repository.\footnote{\url{https://github.com/HERA-Team/h1c_idr3_validation_notebooks}} See \citet{HERA2022b} for data availability of the per-epoch bandpowers and associated statistical uncertainties.

\bibliographystyle{mnras}
\bibliography{main}

\appendix

\section{Mutual information and entropy}
\label{app:entropy}

In this appendix, we provide a brief review of mutual information and entropy. For a thorough and accessible introduction to information theory, and particularly these concepts, see \citet{Cover} and \citet{MacKay2002}. 

\begin{figure}
    \centering
    \includegraphics[scale=0.62]{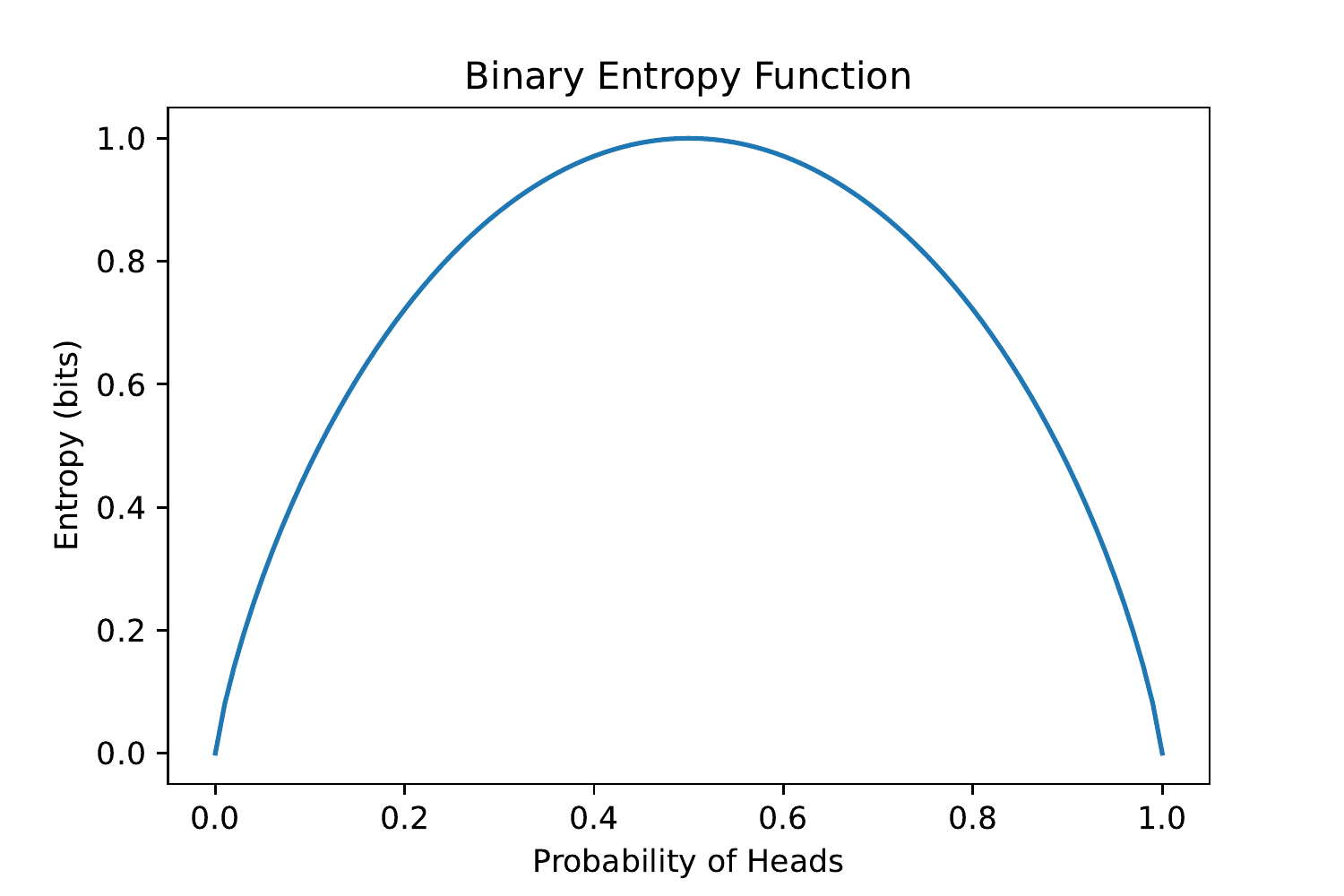}
    \caption{The binary entropy function for a coin toss of varying fairness. The entropy is maximal for a fair coin, and equal to 0 for a coin that always returns one of heads or tails when tossed. Coin flips with highly uncertain outcomes have a high entropy, and inversely, coin flips with highly certain outcomes have low entropy. In general, we can use the entropy as a measure of uncertainty of a random variable.}
    \label{fig:bin_ent}
\end{figure}

We define the Shannon entropy, $S$, of a discrete random variable, $X$, as
\begin{equation}
    S(X) = -\sum_x P(x) \log_2(P(x))
\end{equation}
where $P(x)$ is its probability mass function and the sum is over all states obtainable by the random variable. A similar definition exists for continuous random variables, and is referred to as ``differential entropy." This is interpreted as a measure of uncertainty for the random variable, and is measured in bits if $\log_2$ is used. To understand this interpretation, consider a 2-sided coin with probability $p$ of returning a particular side (say ``heads") after a toss. The entropy of the coin is
\begin{equation}
    S_2(p) \equiv -p\log_2(p) - (1 - p)\log_2(1-p) ,
    \label{eq:bin_ent}
\end{equation}
which is sometimes referred to as the binary entropy function. This function is plotted in Figure \ref{fig:bin_ent}. It is maximal for $p=1/2$, expressing the fact that the outcome of a fair coin is minimally predictable (or maximally uncertain) compared to any unfair one. In the limit as $p$ goes to 0 or 1, the binary entropy function goes to 0, expressing that such a coin will always occupy a certain state after a toss. 

We define the conditional entropy of the random variable, $X$, conditioned on another random variable, $Y$, as
\begin{equation}
    S(X|Y) \equiv -\sum_{x,y}P(x,y)\log_2(P(x | y))
\end{equation}
where $P(x,y)$ is the joint probability mass function of $X$ and $Y$, and $P(x|y)$ is the conditional distribution of $X$ given $Y$. This should be interpreted as the average remaining uncertainty about $X$ after considering a known value for $Y$. When $X$ and $Y$ are independent, this quantity is equal to $S(X)$, and when the state of $Y$ completely specifies the state of $X$, this quantity is equal to 0. For our purposes, we can define the mutual information between $X$ and $Y$ as
\begin{equation}
\begin{aligned}
        I(X;Y) & \equiv S(X) - S(X | Y) \\
        & = S(Y) - S(Y | X)
        \label{eq:mut_info_def}
\end{aligned}
\end{equation}
where the second equality indicates symmetry in the argument. In concordance with our previously stated interpretations, the mutual information is the average reduction in uncertainty about one random variable after gaining knowledge of the state of the other. If (and only if) $X$ and $Y$ are independent, they have 0 mutual information, and therefore knowledge about one cannot reduce uncertainty about the state of the other. The mutual information is always non-negative \citep{Cover}. 

In an inference problem, we are concerned with how informative the data is for a given set of hypotheses. In other words, we want to know $I(\HH; \bold{d})$, where $\HH$ is the set of hypotheses. Due to $P(\bold{d})$ being a mixture distribution, this problem prohibits analytic calculation of the mutual information. However, we can quickly estimate it using Monte Carlo methods using the second form of Equation \ref{eq:mut_info_def}. In this case, we have\footnote{For brevity, we are dropping $\bold{C}_0$ in this expression.} 
\begin{equation}
    I(\HH;\bold{d}) = S(\bold{d}) - \sum_{i=0}^{Q-1}P(\HH_i)S(\bold{d} | \HH_i).
\end{equation}
Since the conditional distributions are all Gaussian with known mean and covariance, their individual entropies can be calculated analytically. The left-hand term involves integrating the logarithm of a Gaussian mixture against that mixture. To estimate this term, we generate $D$ data points from the mixture distribution $P(\bold{d}) = \sum_i P(\bold{d} | \HH_i)P(\HH_i)$, and then compute the sample entropy:
\begin{equation}
    \hat{S}(\bold{d}) \equiv -\frac{1}{D}\sum_{j = 1}^D \log_2\bigg(\sum_{i=0}^{Q-1}P(\bold{d}^{(j)} | \HH_i)P(\HH_i)\bigg)
\end{equation}
where $\bold{d}^{(j)}$ is the $j$th sample of the mixture. The mutual information between a hypothetical data set and the hypothesis may be estimated as
\begin{equation}
    \hat{I}(\mathcal{H}; \bold{d}) = \hat{S}(\bold{d}) - \sum_{i=0}^{Q-1}\frac{1}{2}\log_2(\det(2\pi e \bold{C}'_i))P(\HH_i).
\end{equation}

\section{Evaluating the Posterior for Vector Valued Measurements}
\label{app:multi_post}

In section \ref{sec:post_eval}, we showed how to evaluate the posterior probability for a situation in which multiple measurements were made of a single quantity, such as several measurements of a single wave mode in the 21cm power spectrum. However, it is sometimes of interest to combine information about several measured quantities, such as examining several modes in the power spectrum simultaneously. This will be useful in HERA since systematics can be mildly delocalized (cable reflections; \citet{Kern2020a} or totally delocalized (RFI; \citet{Wilensky2020} in the power spectrum. The general result of this consideration is that one takes the result from \S\ref{sec:post_eval} and evaluates all elements of vector and matrix quantities blockwise. In more detail, let $\bold{d}_j$ be a column vector of length $M$ representing the $j$th measurement of the $M$ quantities (wave modes) in consideration. Then let\footnote{The exact packing order is inconsequential as long as it is consistent between quantites, but this is a natural packing order for the HERA problem since epochs are uncorrelated, meaning this packing produces a block-diagonal $\bold{C}_0$.}
\begin{equation}
    \bold{d} = \begin{pmatrix}
                \bold{d}_0 \\
                \bold{d}_1 \\
                \vdots \\
                \bold{d}_{N-1}
               \end{pmatrix}
\end{equation}  
and define
\begin{equation}
    \boldsymbol{\mu} \equiv \begin{pmatrix}
    \mu_0 \\
    \mu_1 \\
    \vdots \\
    \mu_{M-1}
    \end{pmatrix}
\end{equation}      
to hold the expected values of the separate quantities (power in each wave mode) under the null hypothesis. We then pack $\eps$ in the same way as $\bold{d}$, and each element of $\bold{C}_0$ and $\bold{C}_{\eps, i}$ is replaced by a matrix of shape $M\times M$. For the HERA example, each diagonal block represents the covariance between the different wave-modes within a given epoch, and the off-diagonal blocks describe correlations between both wave modes and different epochs. 

Since the common mean is now vector valued, the final marginalization is over the vector, $\boldsymbol{\mu}$. This means that in the case of $\HH_i$, the marginalization over the biases must also report something over that vector. The correct answer is to do the following replacement:
\begin{equation}
    \mu_0\mathds{1} \rightarrow \mathds{1}\otimes\boldsymbol{\mu}.
\end{equation}
The symbol, $\otimes$, represents the Kroneckcer product. This replacement takes $\boldsymbol{\mu}$ and concatenates $N$ copies of it columnwise. We then call the result of the bias marginalization $f(\boldsymbol{\mu})$, which is now an improperly normalized multivariate Gaussian over $\boldsymbol{\mu}$. Proceeding as in the previous section, one may see that this Gaussian has covariance matrix, $\tilde{\bold{C}}_i$, equal to
\begin{equation}
   \tilde{\bold{C}}_i^{-1} = \sum_{m=0}^{N-1}\sum_{n=0}^{N-1}(\bold{C}_0 + \bold{C}_{\eps, i})^{-1}_{mn}
\end{equation}
where each element of the sum is the $mn$'th block of shape $M\times M$ in the matrix $(\bold{C}_0 + \bold{C}_{\eps, i})^{-1}$. This is really just the blockwise generalization of Equation \ref{eq:sig_tild}, where in that case the operation sums over scalar elements of the inverse covariance matrix, rather than its blocks. In other words, if one takes Equation \ref{eq:sig_tild} and writes it in terms of elementwise addition rather than in matrix notation, one achieves the identical expression as above but with the meaning of the symbols slightly changed. We can package this using the Kronecker product to obtain
\begin{equation}
   \tilde{\bold{C}}_i^{-1}= (\mathds{1}\otimes\bold{I}_M)^T(\bold{C}_0 + \bold{C}_{\eps, i})^{-1}(\mathds{1}\otimes\bold{I}_M)
\end{equation}
where $\bold{I}_M$ is the $M\times M$ identity matrix.
Similarly, Equation \ref{eq:mu_tild} generalizes to 
\begin{equation}
    \tilde{\boldsymbol{\mu}} = \tilde{\bold{C}}_i(\mathds{1}\otimes\bold{I}_M)^T(\bold{C}_0 + \bold{C}_{\eps, i})^{-1}(\bold{d} - \bmu_{\eps, i}).
\end{equation}
Following all this though, Equations \ref{eq:final_mean} and \ref{eq:final_cov} generalize to
\begin{equation}
    \bmu'_i = \bmu_{\eps,i} + \mathds{1}\otimes\bmu.
\end{equation}
\begin{multline}
    (\bold{C}'_i)^{-1} = (\bold{C}_0 + \bold{C}_{\eps, i})^{-1} - \\  (\bold{C}_0 + \bold{C}_{\eps, i})^{-1}(\mathds{1}\otimes\bold{I}_M)(\bold{C}_p^{-1} + \tilde{\bold{C}}_i^{-1})^{-1}(\mathds{1}\otimes\bold{I}_M)^T(\bold{C}_0 + \bold{C}_{\eps, i})^{-1}
\end{multline}
where we have implicitly generalized the prior over the common mean to a multivariate Gaussian.

For interpretations of $\boldsymbol{\mu}$ where Gaussian priors are inappropriate, the final multivariate marginalization can present significantly more numerical difficulty if many quantities (modes) are considered simultaneously. 

\section{Partitioning a Bias Configuration into Correlated and Uncorrelated Subsets}
\label{sec:app_part}

In most of this paper, we consider the hypothesis set in which any combination of subsets of data may be biased, each datum has one mean value it may take when biased, the bias priors have diagonal covariances, and only data biased under the hypothesis have nonzero covariance entries. Due to these last two points, we refer to this as the ``diagonal hypothesis set." In certain applications, it may be beneficial to ``partition" each diagonal hypothesis into correlated and uncorrelated subsets. As we show, this generates a huge number of hypotheses, however it may be useful for investigating particular systematic effects that are expected to affect some data and not others. 

As a simple example, we consider the $N = 2$ case with equal bias prior widths in all hypotheses. As pointed out in the main text, the null hypothesis corresponds to the bias covariance
\begin{equation}
    \bold{C}_{\eps,0} = \begin{pmatrix}
    0 & 0 \\
    0 & 0
    \end{pmatrix}
\end{equation}
Assuming equal width in the bias priors for the alternative hypotheses, we can write the rest of the diagonal entries down as
\begin{equation}
    \{\bold{C}_{\eps, i}: 1 \leq i \leq 3\} = \bigg\{\begin{pmatrix}
    1 & 0 \\
    0 & 0
    \end{pmatrix}, \begin{pmatrix}
    0 & 0 \\
    0 & 1
    \end{pmatrix}, 
    \begin{pmatrix}
    1 & 0 \\
    0 & 1
    \end{pmatrix}\bigg\}.
\end{equation}
This set of hypotheses provides no constraint on the relative equivalence between the biases. We may consider a fifth hypothesis with covariance, $\bold{C}_{\eps, 4}$, defined by
\begin{equation}
    \bold{C}_{{\eps}, 4} \equiv \begin{pmatrix}
    1 & 1 \\
    1 & 1
    \end{pmatrix}.
\end{equation}
Since this matrix is degenerate, this hypothesis establishes an exact equivalence relation between the two bias parameters of the form
\begin{equation}
    \varepsilon_1 = \varepsilon_0 + b
\end{equation}
where $b$ is a constant in terms of the bias prior means for those data. A slope can be added by using different prior widths for the data, and the equivalence can be softened to a correlation by choosing a correlation coefficient less than 1, thus removing the degeneracy. These are all the hypotheses for the $N=2$ case, since no other hypotheses in this set may be further partitioned. 

Let us now take a brief foray into the $N=3$ case before generalizing to arbitrary $N$. When all three diagonal elements are ``on," we have a set of five matrices to consider (let us denote this set, $R$):
\begin{multline}
    R = \Bigg\{\begin{pmatrix}
    1 & 0 & 0 \\
    0 & 1 & 0 \\
    0 & 0 & 1
    \end{pmatrix},
    \begin{pmatrix}
    1 & 1 & 0 \\
    1 & 1 & 0 \\
    0 & 0 & 1
    \end{pmatrix},
    \begin{pmatrix}
    1 & 0 & 1 \\
    0 & 1 & 0 \\
    1 & 0 & 1
    \end{pmatrix},\\
    \begin{pmatrix}
    1 & 0 & 0 \\
    0 & 1 & 1 \\
    0 & 1 & 1
    \end{pmatrix},
    \begin{pmatrix}
    1 & 1 & 1 \\
    1 & 1 & 1 \\
    1 & 1 & 1
    \end{pmatrix}
    \Bigg\}.
\end{multline}
\balance

These represent the five possible equivalence relations between three numbers, where the diagonal matrix reflects that they are all independent, the middle three matrices represent scenarios where two have an exact equivalence while the third is unrelated, and the final matrix represents a scenario in which all are exactly related. Notice however that we must also partition  all $\binom{3}{2}$ matrices in which only two diagonal elements are ``on," which is described by the previous example (two matrices for each pair -- one with off-diagonals ``on" and one with them ``off" i.e. the diagonal case). Of course, we must also include the remaining $\binom{3}{1}$ hypotheses where only one element is ``on," and the null hypothesis in which none are on.

In other words, the total number of hypotheses for the $N$th case, denoted $B_{N+1}$, is such that
\begin{equation}
    B_{N + 1} = \sum_{k=0}^N\binom{N}{k}B_k
\end{equation}
\begin{equation}
    B_0 = 1,
\end{equation}
where this last relation reflects that there is only one null hypothesis.

This turns out to define a set of well-studied numbers known as Bell numbers \citep{oeisbell}. The $N$th bell number describes how many ways there are to partition a set of $N$ elements, or equally well, the total number of equivalence relations on a set of such a size. They also arise in the moments of the Poisson distribution. For our $N$th case, we must sum over all sub-partitions as well as the final case where all diagonal elements are on, meaning we have $B_{N+1}$ hypotheses in consideration for a jackknife with $N$ data points. 
The Bell numbers increase extremely rapidly. The first few are $1, 1, 2, 5, 15, 52, 203, 877, 4140, \ldots$. 

Due to this extremely rapid increase, we do not recommend an exhaustive search over all of the hypotheses. Rather, we recommend surgically choosing a subset of them where equivalence relations are expected. For instance, if two data are expected to be affected by a similar systematic that is not present in other data within the jackknife, then one might include the partition in which this equivalence relation between only these two is posed (e.g. different experimental methods of constraining the Hubble parameter).

\end{document}